\documentclass[twocolumn]{aastex7}
\usepackage{comment}
\usepackage{xcolor}
\usepackage{commath}
\usepackage{soul}
\usepackage{CJKutf8}
\usepackage[normalem]{ulem}
\usepackage{array}
\usepackage{makecell} 
\usepackage{listings}

\newcommand{\vvk}{\boldsymbol{\upsilon}_{\rm K}}
\newcommand{\vkx}{\upsilon_{\rm K,x}}
\newcommand{\vky}{\upsilon_{\rm K,y}}
\newcommand{\vkz}{\upsilon_{\rm K,z}}
\newcommand{\vrel}{\upsilon_{\rm rel}}
\newcommand{\vvrel}{\boldsymbol{\upsilon}_{\rm rel}}

\newcommand{\uu}{u}
\newcommand{\vuu}{\boldsymbol{u}}
\newcommand{\uux}{u_{\rm x}}
\newcommand{\uuy}{u_{\rm y}}
\newcommand{\uuz}{u_{\rm z}}
\newcommand{\jorb}{J_{\rm orb}}

\def\ms{M_\odot}
\def\jorb{J_\mathrm{orb}}
\def\simgr{\mathrel{\hbox{\rlap{\hbox{\lower4pt\hbox{$\sim$}}}\hbox{$>$}}}}

\def\mso{M_\odot}

\def\kms{\mathrm{\,km\,s}^{-1}}
\def\kpc{{\rm \,Kpc}}

\begin{document}

\begin{CJK*}{UTF8}{gbsn}

\title{Large offsets of double-neutron-star mergers: degeneracies and implications for host associations}

\correspondingauthor{Xiao-Tian Xu}
\email{xxu-tdli@sjtu.edu.cn}

\correspondingauthor{Bin-Bin Zhang}
\email{bbzhang@nju.edu.cn}

\correspondingauthor{Wenjie Wu}
\email{wwu@astro.uni-bonn.de}

\author[orcid=0000-0001-9565-9462]{Xiao-Tian Xu (徐啸天)}
\affiliation{Tsung-Dao Lee Institute, Shanghai Jiao-Tong University, 1 Lisuo Road, Shanghai 201210, People’s Republic of China}
\email{xxu-tdli@sjtu.edu.cn}

\author[orcid=0000-0003-4111-5958]{Bin-Bin Zhang}
\affiliation{School of Astronomy and Space Science, Nanjing University, Nanjing 210093, People{'}s Republic of China}
\affiliation{Key Laboratory of Modern Astronomy and Astrophysics (Nanjing University), Ministry of Education, Nanjing 210023, People's Republic of China}
\email{bbzhang@nju.edu.cn}

\author[]{Wenjie Wu}
\affiliation{ Helmholtz-Institut f\"ur Strahlen- und Kernphysik, Universit\"at Bonn, Nussallee 14-16, D-53115 Bonn, Germany}
\email{wwu@astro.uni-bonn.de}

\author{Suoning Wang}
\affiliation{School of Astronomy and Space Science, Nanjing University, Nanjing 210093, People{'}s Republic of China}
\affiliation{Key Laboratory of Modern Astronomy and Astrophysics (Nanjing University), Ministry of Education, Nanjing 210023, People's Republic of China}
\email{602024260014@smail.nju.edu.cn}

\begin{abstract}
The projected offset of a Gamma-Ray Burst (GRB) from its host galaxy is an important diagnostic of its physical origin, where the host associations are usually evaluated through chance-coincidence probabilities. However, as shown by observations, the host associations given by this approach may be ambiguous in some cases. In this work, we revisit the modelling of the offsets of double-neutron-star (DNS) mergers with controlled numerical experiments using Monte Carlo simulations. To isolate the dynamical effects of the second supernova (SN) from uncertainties in the upstream binary evolution, we parametrise the direct progenitor of DNSs and integrate the post-SN trajectories in galactic potentials using \texttt{galpy}. 
Complementary to previous population synthesis calculations, our Monte-Carlo tool, \texttt{GRB-offset}, provides a lightweight framework to identify sensitivities and expose degeneracies in offset modelling.
With \texttt{GRB-offset}, we have investigated the following physical ingredients, supernova physics (ejected mass and kick velocity), galactic potential (potential shape, galaxy virial mass), DNS birth location, star formation history, and redshift. The offset distributions given by our reference Monte-Carlo simulation peak at around 30 kpc and 1000 kpc for the DNSs bound to and unbound from the host galaxy, which sensitively depends on the considered physical ingredients. A given offset can arise from various combinations of these ingredients, introducing substantial degeneracies when interpreting observed offsets. Therefore, an observed offset alone can hardly constrain the properties of the DNS progenitor and of the host galaxies.
\end{abstract}

\keywords{ 
\uat{Gamma-ray bursts}{629} ---
\uat{Binary stars}{154}  --- 
\uat{Stellar astronomy}{1583}
}

\received{}
\revised{}
\accepted{}
\submitjournal{ApJ}

\section{Introduction\label{sec:intro}}

The traditional classification of GRBs by duration does not uniquely trace their progenitor systems. Events with $T_{90}<2\,{\rm s}$ are usually classified as short GRBs (SGRBs), while those with $T_{90}>2\,{\rm s}$ are classified as long GRBs (LGRBs) \citep{Kouveliotou1993ApJ...413L.101K}. However, the physical classification is based on progenitor type: Type I GRBs are associated with compact-object mergers, whereas Type II GRBs are associated with collapsars \citep{Zhang2009ApJ...703.1696Z}. The joint detection of GW170817 and GRB\,170817A established that at least some SGRBs originate from double-neutron-star (DNS) mergers \citep{Abbott2017PhRvL.119p1101A}, which are also expected to be accompanied by afterglow and kilonova emission \citep{Li1998ApJ...507L..59L,Zhang2006ApJ...642..354Z}.

This duration--progenitor correspondence has been further challenged by long-duration Type I GRBs. GRB\,211211A and GRB\,230307A both have long prompt-emission durations but show evidence for a merger origin, including kilonova-like emission \citep{Yang2022Natur.612..232Y,Troja2022Natur.612..228T,Rastinejad2022Natur.612..223R,Levan2024Natur.626..737L,Sun2025NSRev..12E.401S,Chen2025NatAs...9.1701C}. Several progenitor channels have been proposed for such events, including neutron-star--white-dwarf mergers \citep{Yang2022Natur.612..232Y,Zhong2023ApJ...947L..21Z,Xu2025ApJ...994..264X} and black-hole--neutron-star mergers \citep{Rastinejad2022Natur.612..223R,Sarin10.1093/mnras/stac3441}. Nevertheless, a DNS-merger origin cannot be excluded in the absence of direct gravitational-wave constraints. Additional diagnostics are therefore required to distinguish among possible compact-merger channels.

One useful diagnostic is the projected offset between a GRB and its host galaxy \citep{Bloom2002AJ....123.1111B,Levan2008MNRAS.384..541L,Berger2010ApJ...722.1946B,Salvaterra2010MNRAS.406.1248S}. Observed offsets of SGRBs can reach tens of kpc, and in some cases approach $\sim10$--$100\, {\rm kpc}$ \citep{Fong2022ApJ...940...56F,Nugent2022ApJ...940...57N,skobe2026hostgalaxiesmergerenvironments}, whereas most star formation in galaxies occurs within only a few kpc \citep{Nelson2016ApJ...828...27N,Cano-Diaz2016ApJ...821L..26C}. Such large offsets suggest that the progenitor systems can acquire substantial  momentum kicks before merger. In the isolated-binary formation channel, these velocities are naturally produced by mass ejection and natal kicks during supernova explosions, while the long merger delay time allows the compact binary to travel far from its birth site before producing a GRB \citep{Belczynski2002ApJ...571..394B,Tauris2017,Mandel2022,Marchant2024ARA&A..62...21M,Chattaraj2026ApJ...997...52C}.

Observationally, GRB host galaxies are commonly identified using the probability of chance alignment \citep[e.g.,][]{Bloom2002AJ....123.1111B,Levan2008MNRAS.384..541L,Berger2010ApJ...722.1946B,Fong2022ApJ...940...56F,OConnor2022MNRAS.515.4890O,Gaspari2025A&A...699A.113G}. This method is useful but not sufficient in all cases. It is biased toward bright galaxies, which tend to have low chance-coincidence probabilities, and it can become ambiguous when faint galaxies are also located near the burst position \citep[e.g.,][]{Berger2010ApJ...722.1946B,Dichiara2025arXiv251015867D}. 
Observationally, the recent Einstein Probe event EP250704a/GRB\,250704B is associated with an optical transient, whose redshift ($z=0.66$) is different from that of the nearby galaxy ($z=0.11$) with a low chance-coincidence probability \citep{Li2026arXiv260114137L}. While the nature of the optical transient remains unclear, if it is truly associated with the event, this example indicates that the chance-coincidence method may fail in some cases. GRB\,061201 has two nearby galaxies with offsets about 32\,kpc (G1) and 14\,kpc (G2) \citep{Fong2013ApJ...776...18F}. While G1 has a lower chance-coincidence probability, G2 appears to be more likely to be the host of GRB\,061201 \citep{Mao2026arXiv260602032M}. However, recent JWST observations reveal a third candidate host galaxy of $z>1$ near the event (G3), which appear to have a chance-coincidence probability similar to G1 \citep{Troja2026arXiv260701337T}.

These recent observations call for a theoretical reassessment of the factors affecting the offsets of GRBs. 
Theoretical modelling of compact-binary merger offsets has been performed for a long time. Previous studies have primarily compared predicted offset distributions with observed events to test compact-binary merger scenarios and constrain their progenitor properties \citep[e.g.,][]{Belczynski2002ApJ...571..394B, Church2011MNRAS.413.2004C, Behroozi2013ApJ...770...57B,  Gaspari2024A&A...692A..21G, Gaspari2025A&A...699A.113G}, demonstrating the importance of modelling both the progenitor population and the galactic environment. 
However, the predicted offsets can sensitively depend on various uncertain physical ingredients, including host-galaxy properties, adopted supernova physics, and mass transfer physics adopted for binary evolution. 
Complementary to previous works, we aim to  perform a controlled numerical experiment using Monte Carlo simulations to explore how 
individual uncertain ingredients affect the outcome offset distributions. 
Rather than modelling the full binary evolution from the zero-age main sequence, we parametrise the immediate progenitors of the second supernova and focus on the subsequent dynamical evolution.
To achieve this, we construct a lightweight Monte-Carlo-based tool tailored for the study of the dynamical effects of the second supernova event, \texttt{GRB-offset}.

This paper is organized as follows. In Sect.\,\ref{sec:method}, we describe the framework and the input physics for the \texttt{GRB-offset} code. In Sect.\,\ref{sec:result}, we present the predicted offset distributions under different model assumptions. We discuss the implications in Sect.\,\ref{sec:discuss} and 
summarise our work in Sect.\,\ref{sec:conclusion}.

\begin{figure*}
    \centering
    \includegraphics[trim=0 0.8cm 0 0,clip,width=0.9\linewidth]{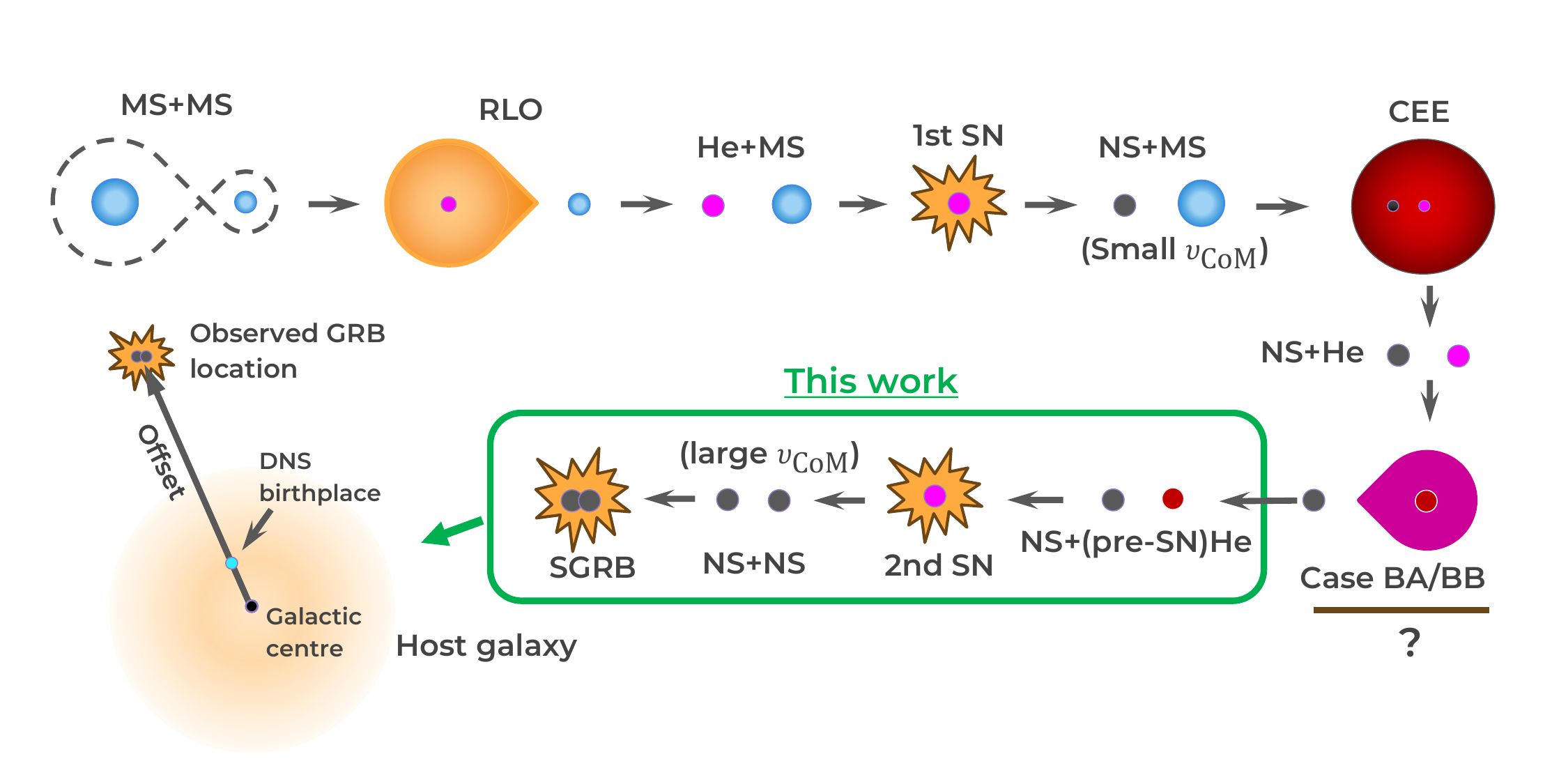}
    \caption{Schematic evolutionary sequence for the formation of double neutron stars (DNS) and the associated offsets at the short $\gamma$-ray burst (SGRB) events. The meaning of the abbreviations are: (1) ``MS", main-sequence star; (2) ``RLO", Roche-lobe overflow; (3) ``He", helium star, or stripped star; (4) ``1st SN", the first supernova event; (5) ``NS", neutron star; (6) ``CEE", common envelope evolution; (7) ``Case BA/BB", the mass transfer with a helium-burning/-depleted star, which may occur before the second SN, reducing the mass of the He star; (8) ``2nd SN", the second supernova event. This work focuses on the dynamical effect of the second SN, as the effect of the 1st SN is ignorable for our purpose (cf., Sect.\,\ref{dis:1st-SN}).}
    \label{fig:schematic}
\end{figure*}

\section{Framework and Method\label{sec:method}}

Figure\,\ref{fig:schematic} presents a schematic evolutionary sequence for the formation of DNSs from isolated binaries, beginning with a main-sequence (MS) binary \citep{Tauris2017,Marchant2024ARA&A..62...21M}. As the primary star evolves, Roche-lobe overflow occurs, producing a helium-star (He)+MS binary, while, for more massive systems, Roche-lobe overflow might be prevented by the luminous blue variable stage \citep{Xu2026arXiv260314840X}. If the mass of the He star exceeds about $3\,\mso$ \citep{Xu2025A&A...704A.218X}, it can end up in a supernova (SN) explosion. If binary disruption is avoided, the system can be observed as a Be X-ray binary \citep{Reig2011}, exhibiting various X-ray outbursts \citep[e.g.,][]{Xu2019,Weng2026SCPMA..6929512W}. The second mass transfer is expected to be unstable because of the extreme mass ratio of the system \citep{Ge2020}, leading to common-envelope evolution \citep{Ivanova2013,Ropke2023LRCA....9....2R}. The system then evolves into a close NS+He binary, and the He star may also fill its Roche lobe \citep{Dewi2002MNRAS.331.1027D,Wang2017MNRAS.472.1593W}. After surviving the second SN event, a DNS is formed, and gravitational-wave radiation drives further orbital shrinkage until the final merger. The corresponding merger delay time can be on the order of Gyrs, allowing the DNS to travel a long distance before producing GRB events.

Binary evolution involves various uncertainties, including mass transfer efficiency and the conditions of surviving the common envelope evolution, 
which significantly affect the resulting DNS population \citep[e.g.,][]{Vigna-Gomez2018}. Our calculation focuses on the dynamical effects of the second SN events outlined in Fig.\,\ref{fig:schematic},
allowing us to treat the properties of the immediate pre-SN binaries parametrically and to isolate the dynamical effects associated with the second SN from the uncertainties in the upstream binary evolution.
The dynamical effects of the first SN event is negligible for our purpose (Sect.\,\ref{dis:1st-SN}).
While we do not sample binaries from the zero-age main sequence and follow their full evolution, 
the code constructed in this work provides lightweight framework for controlled studies of DNS merger offsets, which is publicly accessible\footnote{https://zenodo.org/records/22765322} \citep{xu_2026_22765322}.
In the following, we describe default treatment of the binary and supernova physics, galactic potential, and star formation history.

\subsection{Supernova kick\label{sec:sn-kick}}

In consistent with previous works \citep{Hills1983ApJ...267..322H,Pfahl2002ApJ...573..283P,Hurley2002}, we treat supernova explosion as a sudden mass ejection accompanied with a kick velocity, $\boldsymbol{\upsilon}_{\rm K}$. 

The direct progenitor of a DNS is expected to be the survivor of the common envelope evolution \citep[e.g.,][]{Tauris2017,Marchant2024ARA&A..62...21M,Xu2025ApJ...994..264X,Chattaraj2026ApJ...997...52C}, during which efficient orbital circularisation occurs \citep{Ivanova2013,Ropke2023LRCA....9....2R}.
We accordingly consider a circular pre-SN orbit with a semi-major axis of $a_0$, containing stellar components with masses of $M_1$ and $M_2$. The orbital energy is given by
\begin{equation}
    \frac{1}{2}\vrel^2-\frac{GM_{\rm b,0}}{a_0}= - \frac{GM_{\rm b,0}}{2a_0},
\end{equation}
where $G$ is the gravitational constant, $M_{\rm b,0}$ is the pre-SN binary total mass (i.e., $M_1+M_2$), and $\upsilon_{\rm rel}$ is the relative velocity, 
\begin{equation}
    \upsilon_{\rm rel} = \sqrt{\frac{GM_{\rm b,0}}{a_0}} = \left(\frac{GM_{\rm b,0}}{P_{\rm orb,0}}2\pi\right)^{1/3},
    \label{eq:vper0}
\end{equation}
where $P_{\rm orb,0}$ is the pre-SN orbital period.

Let $M_{\rm 1}$ be the exploding star, leaving a remnant with a mass of $m_1$, which receives a momentum kick of $\boldsymbol{\upsilon}_{\rm K}$. To describe the orientation of kick velocities, we adopt a right-handed coordinate system originating at $M_2$ \citep[Fig.\,A.1 in][]{Hurley2002}. The X-axis has the opposite direction as $\boldsymbol{\upsilon}_{\rm rel}$, and the Y-axis points from $M_2$ to $M_1$. In this coordinate system, we have 
\begin{equation}
    \frac{\boldsymbol{\upsilon}_{\rm rel}}{\upsilon_{\rm rel}} =
        \begin{bmatrix}
-1\\0\\0 
\end{bmatrix}.
\end{equation}
After the SN, the relative velocity, $\boldsymbol{u}$, is given by
\begin{equation}
\begin{split}
    \vuu&= \vvrel + \vvk=
    \begin{bmatrix}
        -\vrel+\vkx\\
        \vky\\
        \vkz
    \end{bmatrix}=
    \begin{bmatrix}
        \uux\\
        \uuy\\
        \uuz
    \end{bmatrix}\\
\end{split},
\end{equation}
where the subscripts, x, y, and z, indicate the components on the X-, Y-, and Z-axis. The post-SN orbital energy, $E_{\rm orb}$, is
\begin{equation}
    \frac{E_{\rm orb}}{\mu}= \frac{1}{2}\uu^2 - \frac{G(m_1+M_2)}{r} = -\frac{G(m_1+M_2)}{2a_{\rm SN}},
    \label{eq:Eorb}
\end{equation}
where $\mu$ is the reduced mass, $m_1M_2/(m_1+M_2)$, and $a_{\rm SN}$ is the semi-major axis of the post-SN orbit. The binary system can survive the SN if $E_{\rm orb} < 0$. According to the conservation of momentum, the SN explosion also generate a momentum kick on the centre of mass, $\boldsymbol{\upsilon}_{\rm CoM}$, which is
\begin{equation}
   (m_1+M_2) \boldsymbol{\upsilon}_{\rm CoM} = m_1 (\boldsymbol{\upsilon}_1+\vvk) + M_2 \boldsymbol{\upsilon}_2,
\end{equation}i.e.,
\begin{equation}
    \boldsymbol{\upsilon}_{\rm CoM} = \frac{m_1}{m_1+M_2}\vvk -
    \frac{M_1 - m_1}{m_1+M_2}\frac{M_2}{M_1+M_2}\vvrel.
    \label{eq:vcom}
\end{equation}
The eccentricity, $e_{\rm SN}$, of the post-SN orbit can be solved by considering the angular momentum, $J_{\rm orb}$, which is
\begin{equation}
    \jorb=\mu |\boldsymbol{r} \times\boldsymbol{u}|
    =\mu\sqrt{G(m_1+M_2)a_{\rm SN}(1-e_{\rm SN}^2)},
\end{equation}
where $\boldsymbol{r}=r(0,1,0)$, resulting in
\begin{equation}
    e_{\rm SN}^2=1-\frac{a_0^2(\uuz^2+\uux^2)}{G(m_1+M_2)a_{\rm SN}},
\end{equation}
or equivalently,
\begin{equation}
    e_{\rm SN}^2 = 1+\frac{2J_{\rm orb}^2E_{\rm orb}}{G^2(m_1+M_2)^2\mu^3}.
\label{eq:ecc}
\end{equation}
The orbital period of the post-SN orbit is given by the Kepler's 3rd law,
\begin{equation}
    P_{\rm orb,SN} = \sqrt{\frac{a_{\rm SN}^34\pi^2}{G(m_1+M_2)}}.
    \label{eq:porb}
\end{equation}

\subsection{Galactic potential and offset}

We treat the DNS progenitor as a test particle in its host galaxy, located at a galactocentric distance of $r_{\rm birth}$. We set the pre-SN space velocity, $\upsilon_{\rm ini,gal}$, of the DNS progenitor to be the rotational velocity at its birth place given by the adopted potential.
After the SN event, the newborn DNS system receives a space velocity of $\boldsymbol{\upsilon}_{\rm CoM}$, which we assume to be randomly orientated.
Similar to Sect.\,\ref{sec:sn-kick}, we consider a right-handed coordinate system, originated at the centre of the host galaxy with the X-axis oppositely oriented to $\boldsymbol{\upsilon}_{\rm ini,gal}$. Then, the space velocity of the newborn DNS, $\boldsymbol{u}_{\rm gal}$, is given by 
\begin{equation}
\begin{split}
    \boldsymbol{u}_{\rm gal}&= \boldsymbol{\upsilon}_{\rm ini,gal} +\boldsymbol{\upsilon}_{\rm CoM}\\
    &=
    \begin{bmatrix}
        -\upsilon_{\rm ini,gal}+\upsilon_{\rm CoM,x}\\
        \upsilon_{\rm CoM,y}\\
        \upsilon_{\rm CoM,z}
    \end{bmatrix}=
    \begin{bmatrix}
        u_{\rm gal,x}\\
        u_{\rm gal,y}\\
        u_{\rm gal,z}
    \end{bmatrix}\\
\end{split},
\end{equation}
where $\boldsymbol{\upsilon}_{\rm CoM}$ is assumed to be randomly orientated with respect to the galactic potential.

The \texttt{GRB-offset} code provides four types of galactic potential by default, isothermal potential, pseudo-isothermal potential, Milky Way-like potential, and Keplerian potential (see Sect.\,\ref{app:isothermal} for details), which can be customised. We adopt the isothermal potential in our reference model, and 
we integrate the trajectory by using \texttt{galpy} \citep{Bovy2015ApJS..216...29B} for both bound and unbound DNSs, which is an open-source python package for galactic dynamics\footnote{\url{https://docs.galpy.org/en/v1.11.2/}}. Then we determine the offset of the resulting SGRB as the galactocentric distance at merger.

To account for the effect of projection, we assume the direction of the line of sight relative to the host galaxy is randomly orientated, which means that the solid angle $d\Omega=d\cos\phi\,d\theta$ is uniformly distributed, where $\theta$ is the azimuthal angle, and $\phi$ is the polar angle measured from the line of sight. Here random orientation requires $\cos\phi$ is uniformly distributed in $[-1,\,1]$, and the projected offset is given by $r_{\rm proj}=r_{\rm offset}\sin\phi$.

\subsection{Merger delay time and star formation history\label{sec:csfr-tauM}}

\begin{figure}
    \centering
    \includegraphics[width=\linewidth]{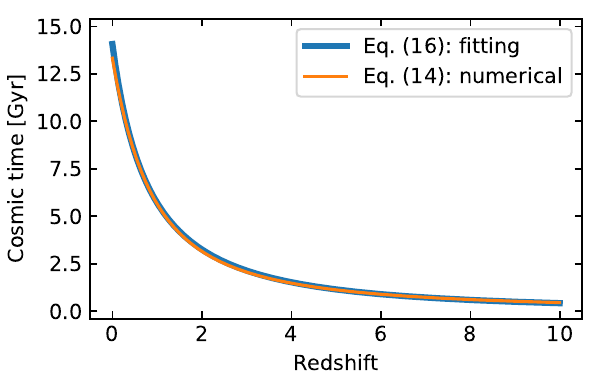}
    \caption{Comparison between the numerical solution of Eq.\,\eqref{eq:t_H} and the fitting formula Eq.\,\eqref{eq:z_fitting}}
    \label{fig:fitting}
\end{figure}

The event rate of DNS mergers at a given redshift, $z_{\rm GRB}$, should be propositional to the star formation rate at the birth redshift, $z_{\rm birth}$, of the DNS progenitors, which depends on the star formation history of the host galaxy. 
We treat the star formation history as one of the uncertain physical ingredient in our numerical experiment, and we vary the star formation histories in Sect.\,\ref{res:param-study}. By default, \texttt{GRB-offset} uses the cosmic star formation rate (SFR) in \citet{Madau2014} as an proxy, which is
\begin{equation}
    \mathrm{SFR}(z)=0.015 \frac{(1+z)^{2.7}}{1+[(1+z)/2.9]^{5.6}} \,\ms\,\mathrm{yr}^{-1}\,\mathrm{Mpc}^{-3},
    \label{eq:sfr}
\end{equation}
where $z$ is the redshift. 
To determine the birth redshift, we use the merger delay time, $\tau_{\rm M}$, as the look-back time, and we adopt the analytical fitting formula by \citet{Mandel2021RNAAS...5..223M} for $\tau_{\rm M}$, which is based on the result by \citet{Peters1964}. We ignore the evolutionary time of massive binaries before the formation of DNS, as it is less than tens of Myrs, ignorable comparing to $\tau_{\rm M}$.

If a GRB is observed at a redshift of $z_{\rm GRB}$, the local age of the Universe, $t_{\rm H}$, is given by 
\begin{equation}
    t_{\rm H}(z) = t_\mathrm{H0}\int_z^\infty\frac{d\,z'}{(1+z') E(z')},
    \label{eq:t_H}
\end{equation}
where $t_\mathrm{H0}$ is the Hubble time, and 
\begin{equation}
    E(z') = \sqrt{\Omega_\mathrm{M}(1+z')^3+\Omega_\mathrm{k}(1+z')^2+\Omega_\lambda},
\end{equation}
where we take $(\Omega_\mathrm{M},\,\Omega_\mathrm{k},\,\Omega_\Lambda)=(0.3,\,0.0,\,0.7)$ according to \citet{Planck2020A&A...641A...6P}. For computational convenience, we adopt the following fitting formula to approximate the  
$t_{\rm H}\text{--}z$ relation given by Eq.\,\eqref{eq:t_H}, which is 
\begin{equation}
    t_{\rm H}(z)=\frac{C_1}{C_2+(C_3+z)^2}\,{\rm Gyr},
    \label{eq:z_fitting}
\end{equation}
where $C_{1-3}$ have values of 70.9, -5.2, and 3.2 respectively \citep[see also][]{Carmeli2006}. We compare the numerical solution of Eq.\,\eqref{eq:t_H} and the fitting formula in Fig.\,\ref{fig:fitting}, which is good enough for the purpose of this work. Then, for a GRB occurring at $z_{\rm GRB}$ with a delay time of $\tau_{\rm M}$, the birth redshift is evaluated by
\begin{equation}
    t_{\rm H}(z_{\rm GRB}) - \tau_{\rm M} = t_{\rm H}(z_{\rm birth}).
\end{equation}

\subsection{Monte Carlo simulation}\label{sec:popsyn}

We use the Monte Carlo method to generate a sample of the direct progenitors of DNSs (i.e., the NS+(pre-SN)He binary in Fig.\,\ref{fig:schematic}). In the following, we summarise our adopted distribution functions (orbital periods and progenitor masses) and input parameters for our Monte Carlo simulations. We note that realistic distribution functions are sensitive to large uncertainties in stellar and binary physics \citep[e.g.,][]{Chattaraj2026ApJ...997...52C}, and we will address this by varying the distribution functions (cf., Sect.\,\ref{res:param-study}).
\begin{itemize}
    \item The NS mass, $M_{\rm NS}$: for simplicity, we adopt the typical value, 1.4$\mso$, for both 1st- and 2nd-born NSs.
    \item The mass of the pre-SN He star, $M_{\rm He}$: the NS/WD boundary at zero-age main sequence is about $9\mso$, corresponding to a He core mass of about $3\mso$ at core carbon depletion \citep[e.g.,][]{Xu2025A&A...704A.218X}.  We accordingly consider a lower limit, $M_{\rm He,low}$, of $3\mso$ for the mass of the pre-SN He star, and the upper limit is taken to be 6.6$\mso$, which is expected to be the NS/BH boundary \citep{Sukhbold2018,Xu2025A&A...704A.218X}. We adopt a mass function of $M_{\rm He}^{-\alpha}$, with $\alpha=2$. 
    This work aims to explore the most probably offset, which is relatively insensitive to the choice of $\alpha$ (Sect.\,\ref{res:param-study}).
    \item The pre-SN orbital period, $P_{\rm orb,pre-SN}$: the Galactic DNSs that have merger delay times below the Hubble time are expected to have a $P_{\rm orb,pre-SN}$ of about 0.1\,d \citep{Tauris2017}. Hence we accordingly adopt a log-normal distribution, 
    \begin{equation}
        f({\log_{10} P}) = \frac{1}{\sigma_{P}\sqrt{2\pi}} \exp\left[-\frac{(\log_{10}P - \mu_{P})^2}{2\sigma_{P}^2}\right],
        \label{eq:lognorm-P}
    \end{equation}
    with a average value, $\mu_{P}$, of -1 (i.e., 0.1\,d), ranging from -2 to 0 (i.e., 0.01\,d to 1\,d), and a standard deviation, $\sigma_{P}$, of 0.3.
    \item The kick velocity of the second SN, $\boldsymbol{\upsilon}_{\rm K}$: we assume a random orientation for $\boldsymbol{\upsilon}_{\rm K}$, and the magnitude is drawn from the Maxwellian velocity distribution, 
    \begin{equation}
        f(\upsilon_{\rm K})=\sqrt{\frac{2}{\pi}}\frac{\upsilon_{\rm K}^2 \exp [-\upsilon_{\rm _{\rm K}}^2/(2\sigma_{\rm K}^2)] }{\sigma_{\rm K}^3},
    \end{equation}
    where $\sigma_{\rm K}$ is taken to be $50\kms$ for our reference model in line with previous works \citep{Tauris2017,Deng2024ApJ...963...80D,Chattaraj2026ApJ...997...52C}.
    \item Galactic potential parameters: the GRB host sample compiled by \citet{Nugent2022ApJ...940...57N} and \citet{Fong2022ApJ...940...56F} suggests a typical stellar mass of $10^{10}\mso$. 
    The galaxy virial mass is generally much larger than the stellar mass, since the dark-matter halo contains most of the mass, with the stellar-to-halo mass ratio changing with galaxy mass and redshift \citep{Behroozi2013ApJ...770...57B}. In our work, we take $M_{\rm vir}=10^{11}\,M_\odot$ as the reference value and explore variations around it.
    For the DNS birthplace, we assume $r_{\rm birth}=3\kpc$, since star formation usually occurs within a few kpc \citep{Nelson2016ApJ...828...27N,Cano-Diaz2016ApJ...821L..26C}. We refer to Sect.\,\ref{app:isothermal} for detailed descriptions on galactic potentials.
\end{itemize}

To predict the distribution function of offsets at a given redshift, $z_{\rm GRB}$, we only consider the DNS population that has merger delay times below the local age of the Universe, and we use the method described in Sect.\,\ref{sec:csfr-tauM} to determine the birth redshifts of the DNS merging at $z_{\rm GRB}$. The contribution of each DNS is weighted by the star formation rate density at its birth redshift, which is
\begin{equation}
    \text{Fraction}(z_{\rm GRB},k) = \frac{\text{SFR}(z_{\rm birth,k})}{\sum_{\rm n} \text{SFR}(z_{\rm birth, n}) },
\end{equation}
where $k$ is the index of a specific DNS merging at $z_{\rm GRB}$, and $n$ is the index of all the DNS merging at $z_{\rm GRB}$.

In addition, most of the NS+(pre-SN)He binaries get disrupted by the second SN event or are too wide to merge, and only a small fraction of the survivors can merge within the local age of the Universe, specifically at high redshift. This means that it requires a large Monte Carlo sample to resolve the population of merging DNSs. To ensure a statistical meaningful sample size for the resulting merging DNS,  \texttt{GRB-offset} skips disrupted systems and wide DNSs (i.e., $\tau_{\rm M}$ exceeding the local age of the Universe), and the calculation continues until the number of merging DNSs reaches $10^6$.

\section{Results\label{sec:result}}

In Sect.\,\ref{res:reference}, we present the outcome of the default setup of the \texttt{GRB-offset} code (the Reference model), which is the baseline of our parameter study (Sect.\,\ref{res:param-study}).

\subsection{Reference model\label{res:reference}}

\begin{figure}[!t]
    \centering
    \includegraphics[width=\linewidth]{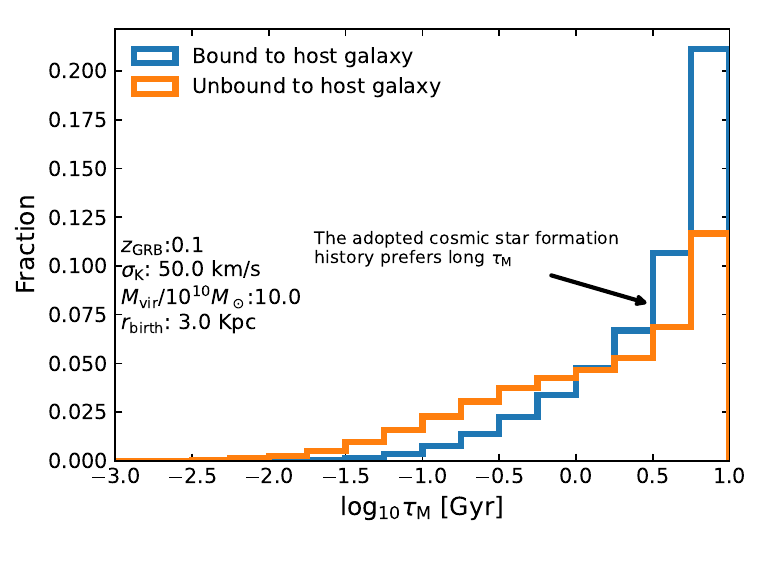}
    \vskip -0.3cm
    \includegraphics[width=\linewidth]{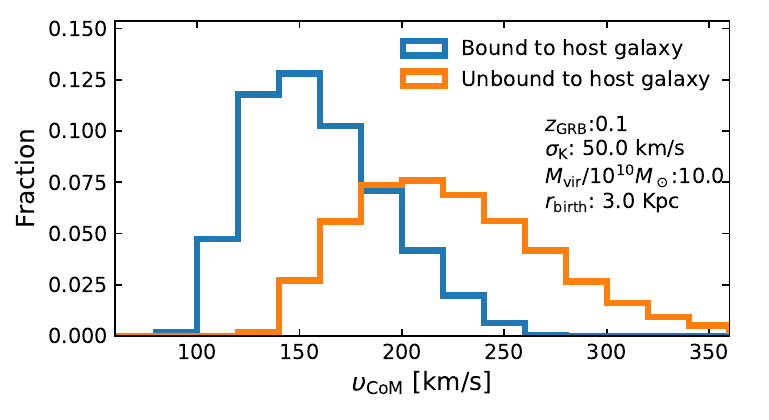}
    \vskip -0.2cm
    \includegraphics[width=\linewidth]{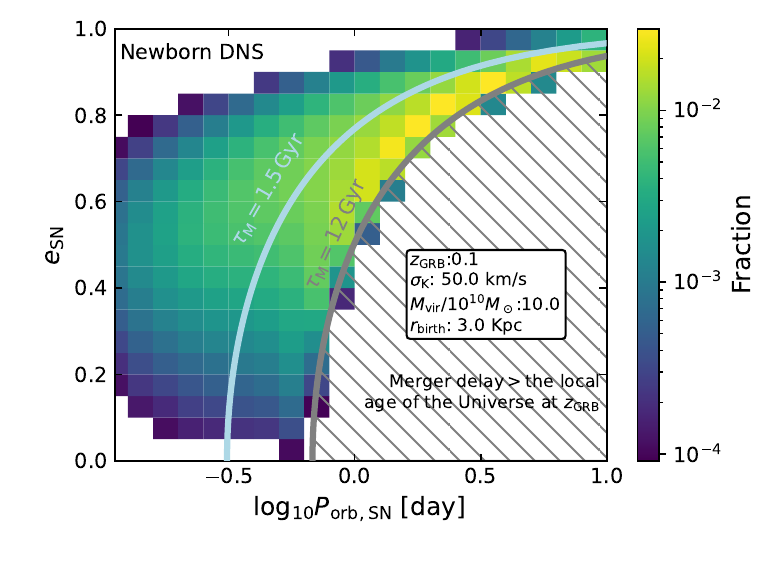}
    \vskip -0.5cm
    \caption{Monte Carlo simulations for the DNS merging at $z_{\rm GRB}=0.1$ by the reference model. The input parameters for the \texttt{GRB-offset} calculation are indicated by the text in each panel. The top and middle panels 
    present the distribution functions of the merger delay time ($\tau_{\rm M}$), and the velocity of the centre of mass ($\upsilon_{\rm CoM}$).
    The blue histograms corresponds to the population bound to the host galaxy, and orange for the unbound population. The bottom panel provides a 2D distribution on the $\log_{10}P_{\rm orb,SN}\text{--}e_{\rm SN}$ plane, with the colormap indicating the predicted fraction in each pixel. The grey hatched region has merger delay time exceeding the local age, 12\,Gyr, of the Universe at $z_{\rm GRB}=0.1$. The lightblue curve marks a merger delay time of 1.5\,Gyr.}
    \label{fig:refmodel-z01-dns}
\end{figure}

\begin{figure}[!t]
    \centering  
    \includegraphics[width=\linewidth]{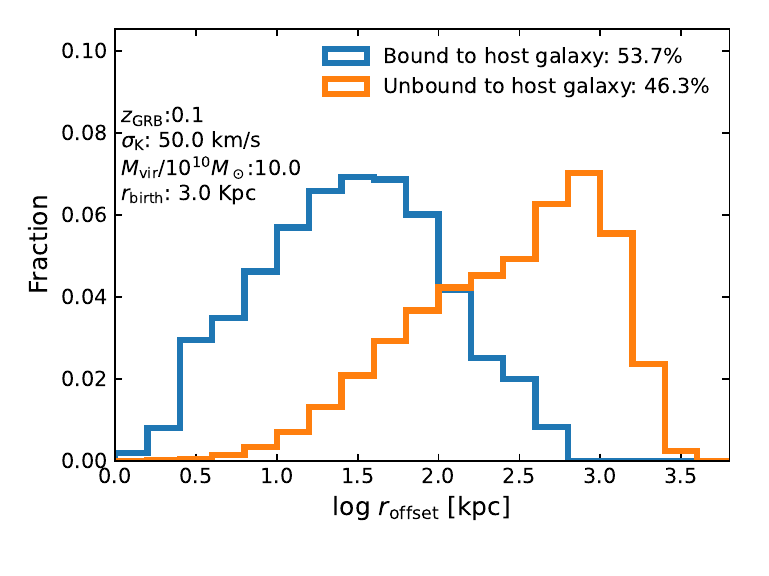}
    \vskip -0.5cm
\includegraphics[width=\linewidth]{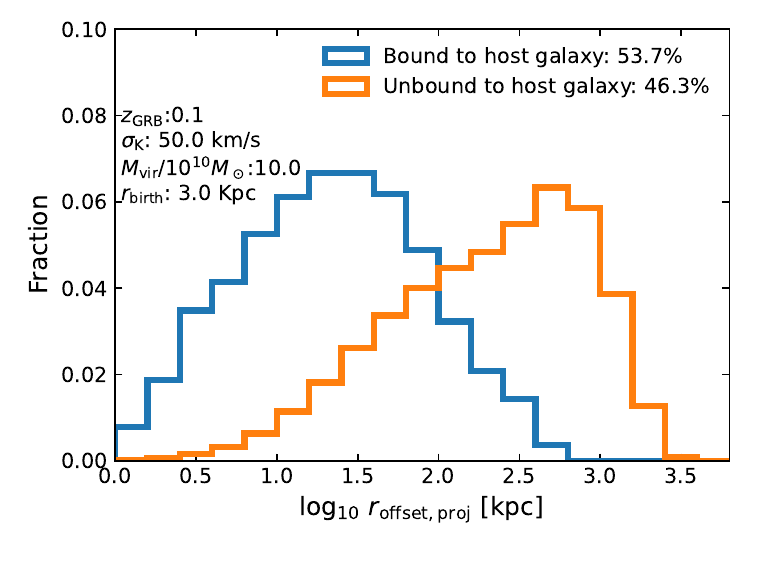}
    \vskip -0.6cm
    \caption{Predicted offset distributions for the DNS merging at $z_{\rm GRB}=0.1$ by the reference model (top panel: without projection, $r_{\rm offset}$; lower panel: with projection, $r_{\rm offset,proj}$). The blue and orange histograms correspond to the population that are bound and unbound to the host galaxy respectively, and the corresponding fractions are indicated in the legend. The text indicates the input parameters for the \texttt{GRB-offset} calculation. 
    }
    \label{fig:refmodel-z01-offset}
\end{figure}

\begin{figure}[!t]
    \centering
    \includegraphics[width=\linewidth]{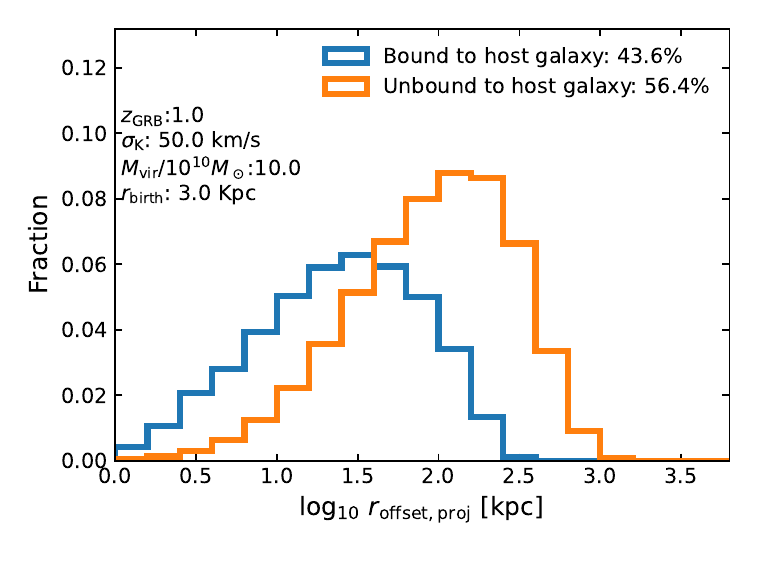}
    \vskip -1cm
    \includegraphics[width=\linewidth]{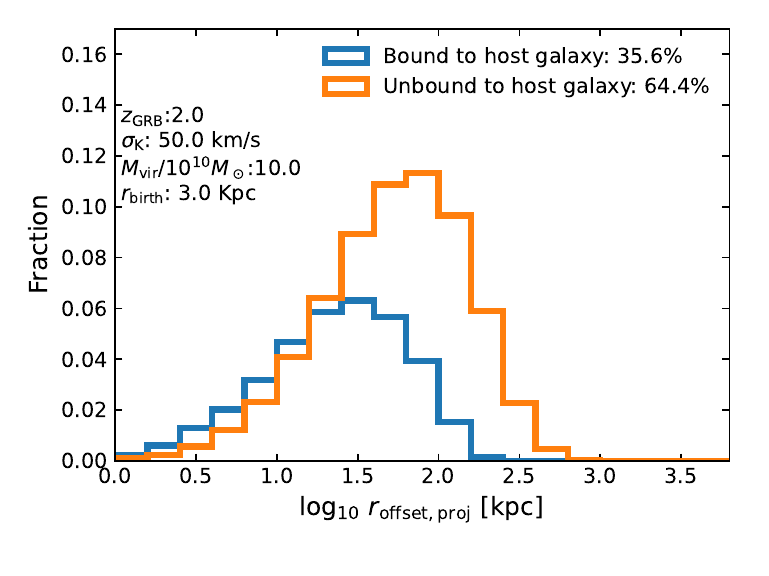}
    \vskip -1cm
    \includegraphics[width=\linewidth]{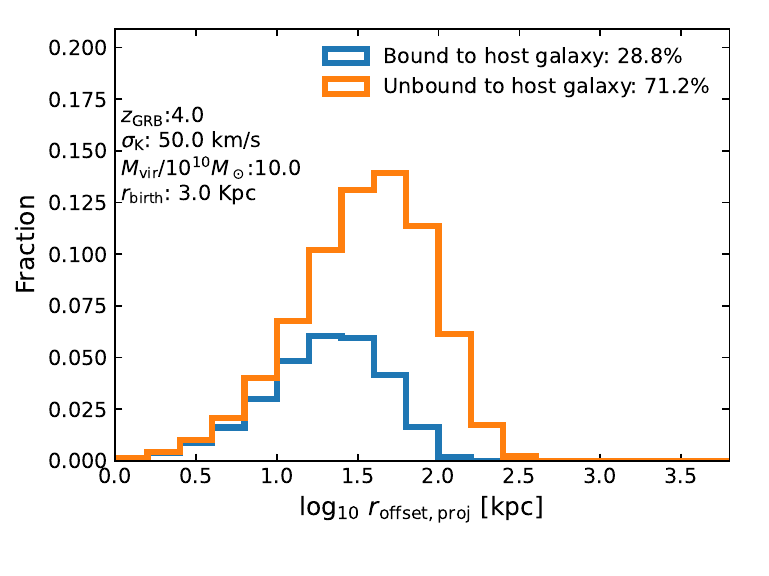}
    \vskip -0.6cm
    \caption{ Projected offset distributions at different redshifts by the reference model (top: $z_{\rm GRB}=1$; middle: $z_{\rm GRB}=2$; bottom: $z_{\rm GRB}=4$). The text and colours have the same meanings as Fig.\,\ref{fig:refmodel-z01-offset}.}
    \label{fig:refmodel-offset-more-z}
\end{figure}

Figure\,\ref{fig:refmodel-z01-dns} presents the properties of the newborn DNS (i.e., right after the second SN) expected to merge at $z_{\rm GRB}=0.1$, where the local age of the Universe is about 12\,Gyr. Unlike previous studies \citep[e.g.,][]{chattaraj2026doubleneutronstardelay}, our expected $\tau_{\rm M}$ piles near the local age of the Universe. 
This is because we consider the cosmic SFR of \citet{Madau2014} for the statistical weighting in the reference model, and we compare it with constant and delayed-$\tau$ star formation histories in Sect.\,\ref{dis:sfr}. 
The shortest merger delay time is about 0.1\,Gyr, roughly corresponding to the merger delay time of a circular DNS with an orbital period of 0.1\,d.

The velocity of the centre of mass, $\upsilon_{\rm CoM}$, of the merging DNSs peaks at about 130$\kms$ and decays to about 350$\kms$, which is mainly determined by the orbital velocity of the progenitor system. For a typical NS+(pre-SN)He binary in our simulation ($3\mso$+1.4$\mso$ in 0.1\,d), the relative orbital velocity is about $750\kms$. For a kick velocity of $50\,\kms$,
the $\upsilon_{\rm K}$ term in Eq.\,\eqref{eq:vcom} is about $25\kms$, while the $\upsilon_{\rm rel}$ term in Eq.\,\eqref{eq:vcom} is about $136\kms$. Therefore, $\upsilon_{\rm CoM}$ increases with increasing ejected mass during SN (i.e., $M_{\rm He}-M_{\rm NS}$) or decreasing pre-SN orbital period (cf., Sect.\,\ref{app:details}), and the fraction drops towards the high-$\upsilon_{\rm CoM}$ regime reflects the decreasing fractions of our input NS+(pre-SN)He binaries below 0.1\,d pre-SN orbital period. 

In addition, we also distinguish the populations that are bound/unbound to the host galaxy. Since the DNSs with high $\upsilon_{\rm CoM}$ have more chance to exceed the escape velocity of the host galaxy, the unbound population has $\upsilon_{\rm CoM}$ (peaking at about $200\kms$) generally higher than that of the bound population (peaking at about $130\kms$), and the unbound population are more likely produced by short pre-SN orbital periods and strong mass ejection during the SN (Sect.\,\ref{app:details}). Meanwhile, the populations have similar $\tau_{\rm M}$ distributions. 

The synthetic merging DNSs exhibit an $e_{\rm SN}\text{--}P_{\rm orb,SN}$ correlation, and most of the merging DNSs have $e_{\rm SN}> 0.5$ and $P_{\rm orb,SN}>1\,$d. The mass ejection and momentum kick can cause orbital widening and eccentricity pumping at the same time \citep[see also][]{Xu2026arXiv260314840X}. Also, for a given $P_{\rm orb,SN}$, $\tau_{\rm M}$ decreases with increasing $e_{\rm SN}$ \citep{Peters1964,Mandel2021RNAAS...5..223M}, which outlines a boundary between merging DNSs and wide DNSs in Fig.\,\ref{fig:refmodel-z01-dns}.

Figure\,\ref{fig:refmodel-z01-offset} presents the offset distribution of the resulting SGRB events at $z_{\rm GRB}=0.1$. 
Fractions of bound (53.6\%) and unbound (46.4\%) systems are comparable. For the DNSs unbound to the host galaxy, the expected offsets peaks near $10^{3}$\,kpc with a maximal offset of about 3000\,kpc. The maximal offset is consistent with the estimate by \citet{Mandel2026ApJ...999L..29M}, which can be understood by the following, 
\begin{equation}
200\kms (\text{typical }\upsilon_{\rm CoM}) \times 10 \text{ Gyr} \simeq 2000\,\text{kpc}.
\end{equation}  
In addition, some unbound DNSs have offsets below 100\,kpc. This is because that their short merger delay times limit the resulting offsets. 
For the bound population, the most probable offset is about 30\,kpc, above which 
the fraction of systems decays to zero at around 300\,kpc. This maximal offset of bound system corresponding to the apastron of a critical orbit, whose periastron is located at $r_{\rm birth}$ and orbital period is twice the local age of the Universe.
Most of the synthetic DNSs have merger delay time near the local age of the Universe, as the systems born at high redshifts have high statistical weights according to the adopted star formation history. 
In addition, we find that the effect of projection does not change the peak value of the distributions.
In the following discussion, we focus on the projected offsets.

Then we present the predicted projected offset distributions at $z_{\rm GRB}=1$, 2, and 4, where the local ages of the Universe are 5.6\,Gyr, 3.2\,Gyr, and 1.5\,Gyr (Fig.\,\ref{fig:refmodel-offset-more-z}). The maximal offsets of both bound and unbound populations are related to the local age of the Universe. For the unbound population, it decreases from about 3000\,kpc ($z_{\rm GRB}=0.1$) to about 300\,kpc ($z_{\rm GRB}=4$). For the bound population, it decreases slightly, from about 300\,kpc ($z_{\rm GRB}=0.1$) to about 100\,kpc ($z_{\rm GRB}=4$), while the peak offsets of the bound population roughly remains unchanged. We note that, for $z_{\rm GRB}>1$, the bound and unbound populations overlap substantially, while the unbound fraction increase with $z_{\rm GRB}$, while the unbound fraction increase from $56.4\%$ ($z_{\rm GRB}=1$) to 71.2\% ($z_{\rm GRB}=4$).

\subsection{Parameter study\label{res:param-study}}

\begin{deluxetable}{c|cccccccc|cc}
     \caption{Definitions of our Monte Carlo models and the fractions of bound/unbound populations}
\tablehead{   &$\sigma_{\rm K}$ & $M_{\rm vir}$ & $r_{\rm birth}$ & $M_{\rm He,low}$ & $\alpha$ & $\mu_{P}$ & $\sigma_{P}$ & Potential& \multicolumn{2}{c}{Fractions}\\     
&$\kms$ & $10^{10}\mso$ & kpc & $\mso$ & &  & 
&&Bound & Unbound
}
 \startdata
    Reference & 50 & $10$ & 3 & 3 & 2 & -1 & 0.3 & iso. &53.7\% & 46.3\% \\
    High-$\sigma_{\rm K}$ & 100 &=&=&=&=&=&=&=&34.6\% & 65.4\% \\
    Low-$\sigma_{\rm K}$ &10 &=&=&=&=&=&=&=&65.0\% & 35.0\% \\ 
    High-$M_{\rm vir}$ &=&30&=&=&=&=&=&=& 90.3\% & 9.7\% \\  
        Low-$M_{\rm vir}$ &=&3&=&=&=&=&=&=&11.1\% & 88.9\% \\
         Lower-$M_{\rm vir}$ &=&1&=&=&=&=&=&=&0.3\% & 99.7\% \\
    Low-$r_{\rm birth}$ &=&=&1&=&=&=&=&=&67.7\% & 32.3\% \\
    High-$r_{\rm birth}$ &=&=&9&=&=&=&=&= &36.3\% & 63.7\% \\
    Low-$M_{\rm He,low}$ &=&=&=&1.5&=&=&=&= &86.8\% & 13.2\% \\
    Flat-$\alpha$ &=&=&=&=&0&=&=&=&47.0\% & 53.0\% \\
    High-$\mu_{P}$ &=&=&=&=&=&0&=&=&79.4\% & 20.6\% \\
    High-$\sigma_{P}$ &=&=&=&=&=&=&1&=&44.3\% & 55.7\% \\
         Pseudo-isothermal&=&=&=&=&=&=&=&pseudo-iso.&51.0\% & 49.0\% \\
         MW-like &=&=&=&=&=&=&=&MW-like &88.3\% & 11.7\% \\
         Keplerian&=&=&=&=&=&=&=&Kep.&87.3\% & 12.7\% \\
\enddata
    \tablecomments{Here ``=" means that the parameters have the same values as the Reference model, and its predictions are presented in Sect.\,\ref{res:reference}. The meanings of the parameters are explained in Sect.\,\ref{sec:popsyn}. For the shape of the potential of the host galaxy, besides the isothermal potential (``iso.") in the reference model, we consider three addition potentials, which are pseudo-isothermal potential (``pseudo-iso."), Milky Way-like potential (``MW-like"), and the Keplerian potential ``Kep.". Here the Keplerian potential approximates the galaxy by a point mass. In the last two columns, we present the fractions of the bound/unbound populations expected by each model.}
    \label{tab:popsyn-model}
\end{deluxetable}

\begin{figure*}[!t]
    \centering
    \includegraphics[trim=0 1.6cm 0 0.4cm, clip,width=\linewidth]{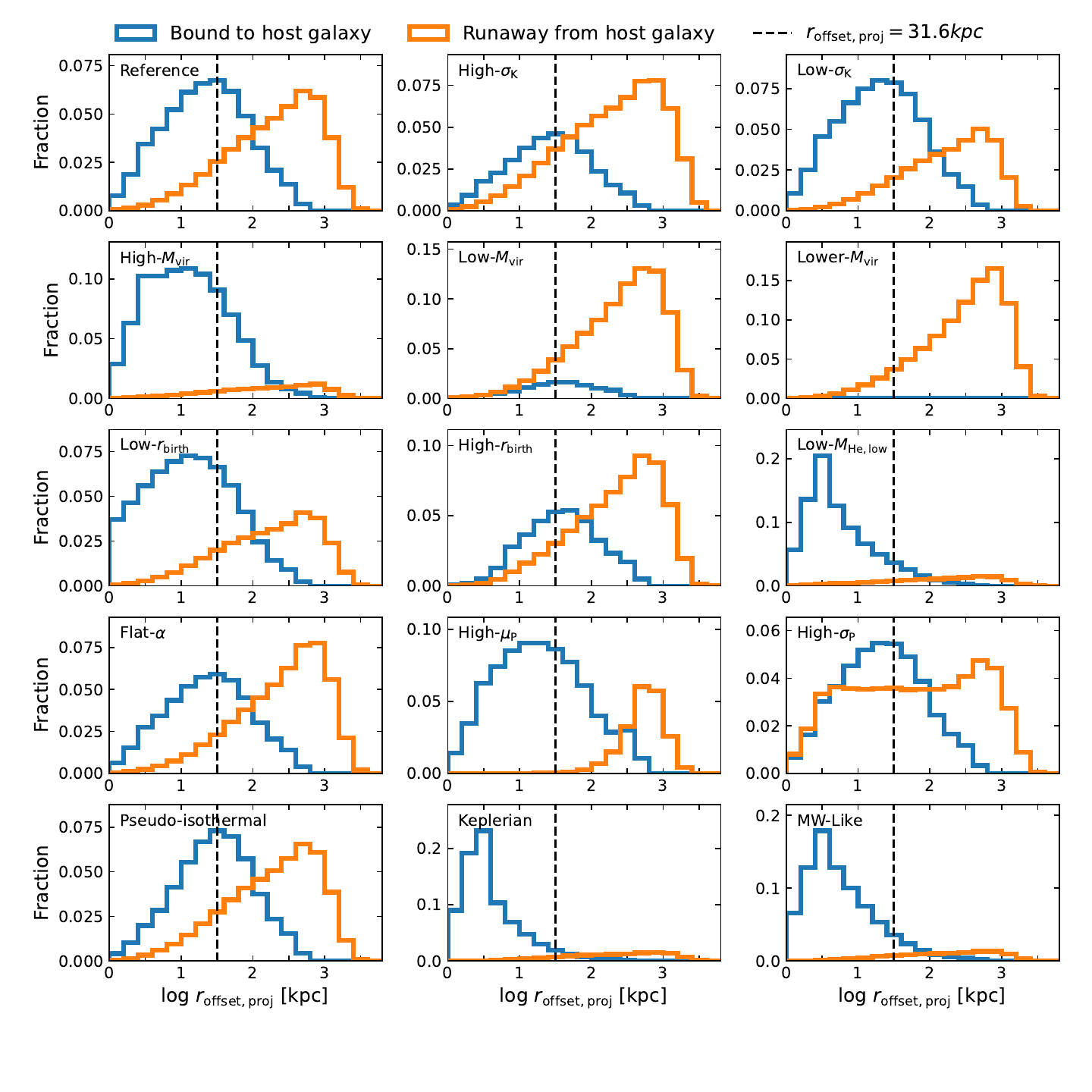}
    \caption{ Projected offsets predicted by different Monte Carlo models (see Tab.\,\ref{tab:popsyn-model} for the input parameters). The identifier of each model is indicated by the upper-left text. The blue and orange histograms correspond to the SGRBs from the DNS populations bound/unbound to the host galaxy. The vertical dashed line marks a offset of 31.6 kpc (1.5 in log).}
    \label{fig:param-study-offset}
\end{figure*}

To better identify the main factors determining the resulting offsets, we perform a parameter study in this section, and the definitions of our Monte Carlo models are summarised in Tab.\,\ref{tab:popsyn-model}, where the Reference model is listed for comparison. Figure\,\ref{fig:param-study-offset} presents the predicted offset distributions at $z_{\rm GRB}=0.1$ by different Monte Carlo models. 
In the following, we describe the results by different Monte Carlo models in detail.

\subsubsection{Supernova kicks}

The properties of supernova kicks is still poorly constrained observationally \citep{Popov2025NewAR.10101734P}. E.g., the bimodality of kick velocity distributions is still under debate \citep{Hobbs2005,Verbunt2017,Disberg2025ApJ...989L...8D}. Meanwhile, growing observational evidence supports an alignment between the spin of NSs and the kick velocities \citep[][and references therein]{Janka2017}. Most recently, it is suggested that an ultra-low SN kick is required to reproduce the population of Galactic Be X-ray binaries \citep{Valli2025arXiv250508857V}. 
We address this uncertainty by taking $\sigma_{\rm K}=100\kms$ (High-$\sigma_{\rm K}$ model) and $10\kms$ (Low-$\sigma_{\rm K}$ model). While changing the magnitude of kick velocities alters the fraction of the bound population (High-$\sigma_{\rm K}$: 34.6\%; Low-$\sigma_{\rm K}$: 65.0\%), the peak offset of the bound population roughly stays unchanged. For the unbound population, the maximal offset is slightly reduced due to the decreasing kick velocity. In addition, the absolute number of the expected merging DNSs should be largely reduced by strong kicks. Investigating this requires sampling binary stars from the zero-age main sequence and including the binary disruption in all SN events, which is beyond the scope of our numerical experiments only focusing on the second SN event.  

\subsubsection{Galactic potential}

Whether a newborn DNS is bound to its host galaxy depending on the properties of the potential \citep{Church2011MNRAS.413.2004C}. We assess these uncertainties by varying the galaxy virial mass,
the location of the birth, and the shape of the galactic potential. The fraction of the bound/unbound population is highly sensitive to the mass of the host. Varying $M_{\rm vir}$ from $3\times10^{11}\mso$ (High-$M_{\rm vir}$ model) to $3\times10^{10}\mso$ (Low-$M_{\rm vir}$ model) and $10^{10}\mso$ (Lower-$M_{\rm vir}$), the bound population deceases from near 100\% to near 0\%. The peak offset is also shifted correspondingly, which is about 10\,kpc in the High-$M_{\rm vir}$ model (Ref. model: $\sim$30\,kpc).  Shifting the birth place from 9\,kpc (High-$r_{\rm birth}$ model) to 1\,kpc (Low-$r_{\rm birth}$ model) also increases the fraction of the bound population from 36.3\% to 67.7\% due to the higher local escape velocity. 

In addition, we examine the offset distributions expected by different types of galactic potentials (See App.\,\ref{app:isothermal} for detailed descriptions of potentials). Comparing with the isothermal potential, the pseudo-isothermal potential assumes a core, which softens the potential, making it slightly less compact than the isothermal potential. Therefore, the Pseudo-isothermal model gives similar offset distributions and bound/unbound fractions to the  Reference model. A Milky Way-like (MW-like) potential is much more centrally concentrated, resulting in an unbound population of 11.7\%, and the peak offsets of the bound population is almost the value of $r_{\rm birth}$. The Keplerian potential, which is the most compact potential, expects a offset distribution concentrated at $r_{\rm birth}$.

\subsubsection{Mass ejection during supernova}

In the Reference model, we have taken 3$\mso$ as the minimal mass of the pre-SN He star, $M_{\rm He,low}$, according to single-star models. However, it has been suggested that the mass of the progenitor of the 2nd-born NS can be further reduced by a Case BA/BB mass transfer \citep{Tauris2015,Wang2017MNRAS.472.1593W,Tauris2017,Chattaraj2026ApJ...997...52C}, resulting in a pre-SN star of $1.5\mso$ and a ultra-stripped SN \citep{Tauris2015}. We accordingly decrease $M_{\rm He,low}$ to $1.5\mso$ in the Low-$M_{\rm He,low}$ model, which drastically alters the resulting offset distribution. The bound population piles up at the birth location, 3 kpc (Ref. Model: $>30\,$kpc). This is because the resulting $\upsilon_{\rm CoM}$ is mainly affected by the $\boldsymbol{\upsilon}_{\rm rel}$ term in Eq.\,\eqref{eq:vcom}. Taking a pre-SN mass of 1.5$\mso$ reduces the $\boldsymbol{\upsilon}_{\rm rel}$ term by a factor of about 10 compared to the case of a $3\mso$ pre-SN star. 

\subsubsection{Input distribution functions}

Our Monte Carlo calculations only focus on the evolution from the second SN event, and we use the Monte Carlo method to populate the NS+(pre-SN)He binaries as our inputs. Here the properties of the NS+(pre-SN)He binaries are affected by the large uncertainties in the common envelope evolution \citep{Ivanova2013,Ropke2023LRCA....9....2R}. To address this uncertainty, we have tested different input distribution functions. 

The Flat-$\alpha$ model considers a flat mass function for the pre-SN He star, which increases the fraction of the unbound population. The second SN events generally eject more material than those in the Reference model, resulting in larger $\upsilon_{\rm CoM}$, and hence larger chances to escape the host galaxy. 
The High-$\mu_{P}$ model shifts the initial logarithmic orbital period function by 1 dex upwards, which reduces the relative velocities in the pre-SN binaries, resulting in lower space velocities. Therefore, this model expects a larger fraction of the bound population. 
The High-$\sigma_{P}$ model uses a flatter orbital period function by increasing $\sigma_{P}$ (Eq.\,\ref{eq:lognorm-P}) to 1. This orbital period function boosts the DNS with low $\tau_{\rm M}$ due to the increasing fraction of close close NS+(pre-SN)He binaries.

\subsubsection{Star formation history\label{dis:sfr}}

\begin{figure*}
    \centering
    \includegraphics[width=0.4\linewidth]{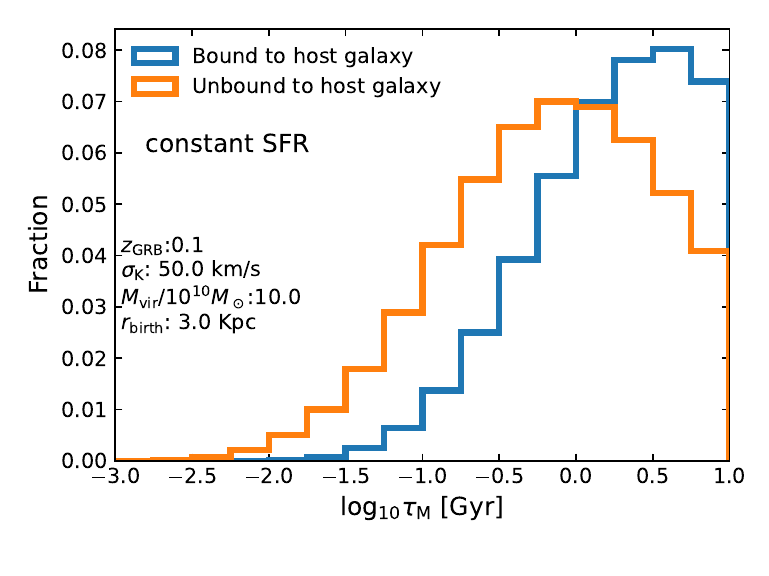}   
    \includegraphics[width=0.4\linewidth]{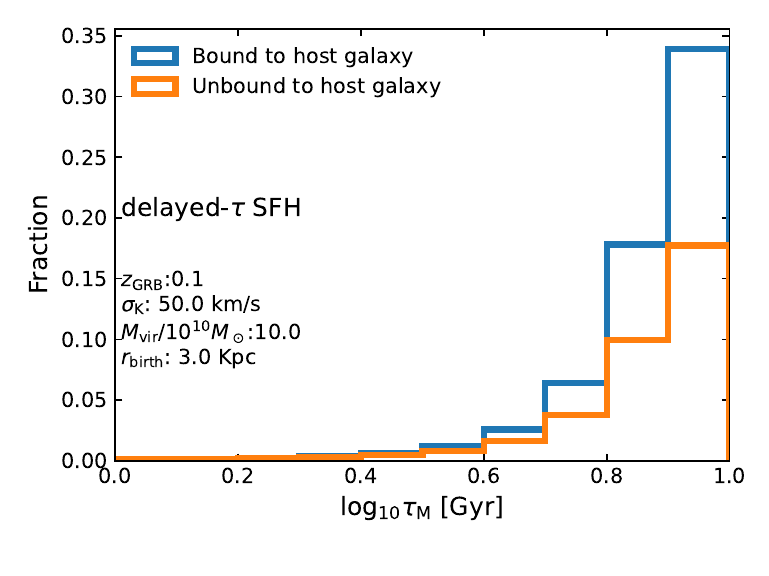}
    \vskip -0.5cm
    \includegraphics[width=0.4\linewidth]{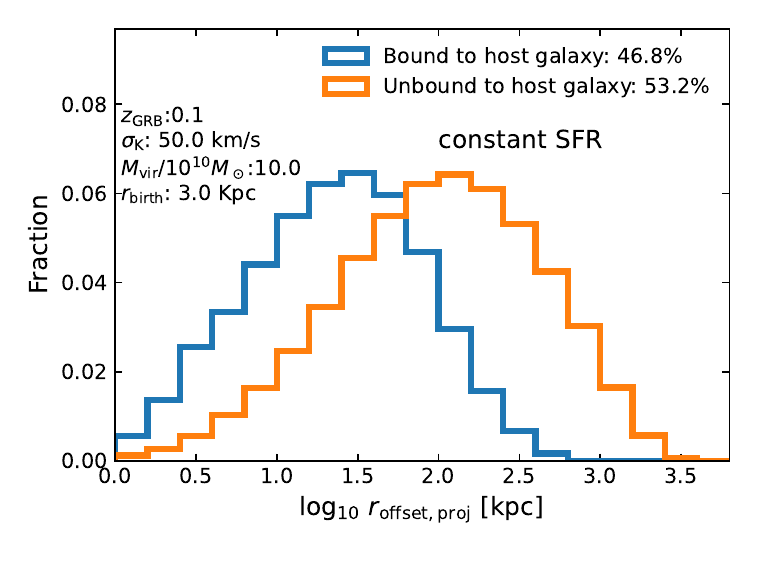}
    \includegraphics[width=0.4\linewidth]{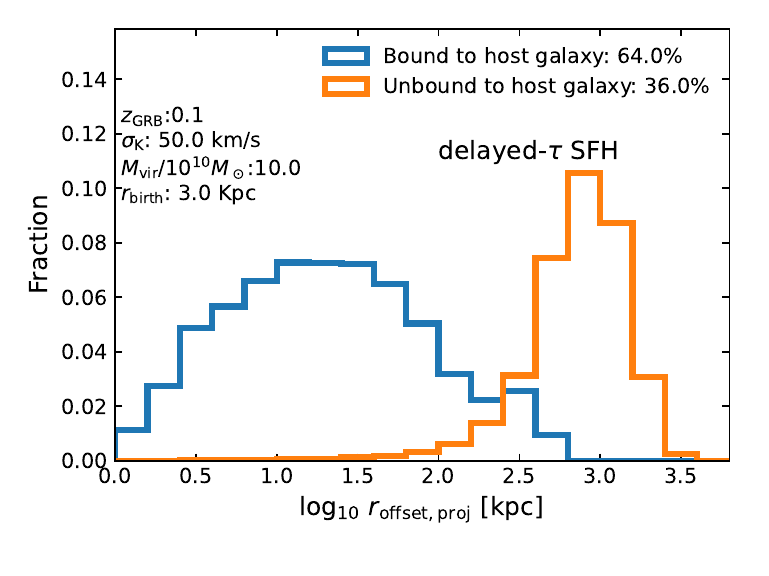}
    \vskip -0.4cm
    \caption{Effects of star formation histories. The left column is computed by a constant star formation rate, and the right column is computed by the delayed-$\tau$ star formation history. The first row present the expected merger time distribution, and the second row present the projected offset distribution. The colours and text have the same meaning as Fig.\,\ref{fig:refmodel-z01-offset}.}
    \label{fig:const-SFR}
\end{figure*}

In our calculations, we have adopted the cosmic star formation rate density in \citet{Madau2014} as the statistical weighting. In this subsection, we discuss the impact of star formation histories by re-computing the reference model with a constant SFR and the delayed-$\tau$ SFH \citep[see also][]{Nugent2022ApJ...940...57N}. Here the delayed-$\tau$ SFH is given by
\begin{equation}
    \text{SFR(z)}\propto \Delta t_{\rm SF}\exp \left(-\frac{\Delta t_{\rm SF}}{\tau}\right),
\end{equation}
where $\tau$ is a constant related to the properties of the galaxy, and $\Delta t_{\rm SF}$ is the difference between the birth time of the DNS  progenitor, $t_{\rm H}(z_{\rm birth})$, and the onset of the star formation, $t_{\rm H}(z_{\rm onset,SF})$, 
\begin{equation}
    \Delta t_{\rm SF} =  t_\mathrm{H}(z_{\rm birth}) - t_{\rm H}(z_{\rm onset,SF}),
\end{equation}
where $z_{\rm onset,SF}$ is the redshift at the onset of the star formation (i.e., SFR($z>z_{\rm onset,SF}$)=0). For this numerical experiment, we take $\tau=1\,\text{Gyr}$ and $t_{\rm H}(z_{\rm onset,SF})=2\,\text{Gyr}$ to illustrate an early-peaked SFH.

Comparing the cosmic SFH adopted in the reference model, the constant star formation expects more DNS formed at low redshift (i.e., short $\tau_{\rm M}$), and the delayed-$\tau$ SFH emphasises those with long $\tau_{\rm M}$. We find that varying SFH can drastically alter the resulting DNS population (Fig.\,\ref{fig:const-SFR}). With a constant star formation rate, $\tau_{\rm M}$ peaks at 3\,Gyr and 0.3\,Gyr for the bound and unbound populations respectively, while in the reference model, the population piles up near the age of the Universe. The population expected by the delayed-$\tau$ SFH piles up at about 10\,Gyr even more than the reference model. For the offset distribution, while the peak of the bound systems remains similar to that by the reference model, the peak of the unbound systems is reduced to about 100\,kpc with a constant star formation rate. The distributions expected by the delayed-$\tau$ SFH resembles that by the reference model. In addition, we also provide a parameter with the constant star formation and the delayed-$\tau$ SFH in App.\,\ref{app:details}.

\section{Discussion\label{sec:discuss}}

\subsection{Implications for interpreting observed offsets}

Our results show that a given GRB offset can arise from different combinations of progenitor and host-galaxy properties. For example, the space velocity of a newborn DNS sensitively depends on both the mass ejected during the second SN and the pre-SN orbital configuration, and the following motion is shaped by the galactic potential, birth location, and star-formation history. In addition, maximal offset is also limited by the local age of the Universe, which may have significant effects for high-redshift GRB events. These combinations of physical ingredients introduce substantial degeneracies when interpreting observed GRB offsets.

Therefore, an observed offset alone can hardly constrain the properties of the DNS progenitor or the likelihood of a proposed host galaxy. In the case of GRB\,061201, the offsets of the nearby galaxies are about 32\,kpc (G1), 14\,kpc (G2), and 1.7$\pm1$\,kpc (G3) \citep{Fong2013ApJ...776...18F,Troja2026arXiv260701337T}. Most of our Monte Carlo simulations produce offset distributions peaking at around 32\,kpc, while producing peak offsets below 10\,kpc generally requires compact potentials, weak mass ejection, or birth place near the galactic centre (Fig.\,\ref{fig:param-study-offset}). However, the peak locations cannot by themselves rank the candidate hosts, as individual events can occur far away from the distribution peak. Therefore, a more robust host galaxy determination requires the offset information to be combined with the chance-coincidence probability and independent constraints on the candidate host.

\subsection{First supernova event\label{dis:1st-SN}}

After the 1st SN event, a binary system can emerge as a Be X-ray binary, whose progenitor in theoretical calculations typically contains a $10\mso$ Be-type companion star and a $4\mso$ He star \citep{Xu2025A&A...704A.218X}. Then the velocity of the centre of mass associated with the 1st SN, $\upsilon_{\rm 1st-SN,CoM}$, is estimated as the following,
\begin{equation}
\begin{split}
\boldsymbol{\upsilon}_{\rm 1st-SN,CoM} &= \frac{1.4}{1.4+10}\boldsymbol{\upsilon}_{\rm K} - \frac{4-1.4}{1.4+10}\frac{10}{10+4}\boldsymbol{\upsilon}_{\rm rel}\\
&\simeq0.12\boldsymbol{\upsilon}_{\rm K} - 0.16\boldsymbol{\upsilon}_{\rm rel}.
\end{split}
\end{equation}
Surviving the CEE phase (see Fig.\,\ref{fig:schematic}) requires the Be X-ray binary to have an orbital period above about 1\,yr \citep{Tauris2017}. Therefore, we consider an orbital period of 1\,yr for the Be+He binary, whose orbital relative velocity is about 70$\kms$. Considering a kick velocity of $100\kms$, the resulting $\upsilon_{\rm 1st-SN,CoM}$ is on the order of $10\kms$. With a typical main-sequence lifetime for the Be star \citep[30\,Myr;][]{Xu2025A&A...704A.218X}, the maximal displacement associated with the 1st SN is below 1\,kpc, which can be ignored compared to the offsets related to the second SN event.

\subsection{Stellar cluster}

A large fraction of massive binary stars is expected to be born in stellar clusters \citep{2004NewAR..48...47K,2003IAUS..212...80Z}, whose potential can reduce the space velocities of newborn DNSs before they entering the galactic potential  \citep[see also][]{2024MNRAS.530.5155W}. In this subsection, we estimate the correction on $\upsilon_{\rm CoM}$ of newborn DNSs related to their natal clusters.

The correction should be treated as an energy subtraction rather than as a
linear decrement in speed. The effective space velocity, $\upsilon_{\rm CoM,out}$, after escape from the natal cluster is roughly
\begin{equation}
v_{\rm CoM,out} = \left(v_{\rm CoM}^2 - v_{\mathrm{esc,cl}}^2\right)^{1/2},
\end{equation}
where  $\upsilon_{\rm esc,cl}$ is the escape velocity of the natal cluster. 
Young massive stellar clusters have typical masses of about $10^4\text{--}10^5\,\mso$ and typical effective
radii of about $R_{\mathrm{eff}} \approx 3$\,pc \citep{2020SSRv..216...69A}, resulting in $v_{\mathrm{esc,cl}}\sim 10\text{--}50\,\kms$, with only extreme cases reaching $\sim100\,\kms$. 
Since $v_{\mathrm{esc,cl}}$ is generally smaller than our predicted $\upsilon_{\rm CoM}$ (cf., Sect.\,\ref{res:reference}), the reduction, $\Delta \upsilon_{\rm CoM}$, can be estimated as 
\begin{equation}
\begin{split}
\frac{\Delta \upsilon_{\rm CoM}}{\upsilon_{\rm CoM}} & = 1 - \frac{\upsilon_{\rm CoM, out}}{\upsilon_{\rm CoM}}\approx \frac{\upsilon_{\mathrm{esc,cl}}^2}{2\upsilon_{\rm CoM}^2}\\
&\sim12.5\% \left(\frac{\upsilon_{\rm esc,cl}}{50\,\kms}\right)^2\left(\frac{\upsilon_{\rm CoM}}{100\,\kms}\right)^{-2}.
\end{split}
\end{equation} 
Therefore, we do not expect the potential of natal clusters to significantly alter our results, while it can slightly enhance the fraction of the bound populations.

\section{Conclusion\label{sec:conclusion}}

The offset of a GRB event is the galactocentric distance between the GRB and its host galaxy, which provides an useful diagnostic of the nature of the event. The observational determination of the host galaxy relies on the chance-coincidence probability method. However, this method is biased towards bright galaxies and might result in ambiguous host determinations, as implied by recent observations \citep{Li2026arXiv260114137L,Mao2026arXiv260602032M,Troja2026arXiv260701337T}.  

Motivated by these recent observations, this work revisits the modelling of GRB offsets with controlled numerical experiments focusing on the dynamical effects of the second SN event. To achieve this, we construct an open-source Monte-Carlo tool, \texttt{GRB-offset}, tailored for our purpose, which samples the direct progenitor system of the second SN event, (pre-SN)He+NS binaries (cf., Fig. \ref{fig:schematic}), computes the mass ejection and kick velocity associated with the SN, and integrates the post-SN trajectory in galactic potentials using \texttt{galpy}.  This treatment allows us to separate the uncertainties in the upstream binary evolution from the dynamical effects of the second SN event. However, we do not consider the systems disrupted during the SN and the DNSs with $\tau_{\rm M}$ exceeding the local age of the Universe, and our calculation cannot give the absolute numbers of merging DNSs or realistic merger rates.

With \texttt{GRB-offset}, we have investigated the effects of the following physical ingredients, the properties of the host galaxy (virial mass, types of potentials, and star formation histories), supernova physics (kick velocities and ejected mass), and the orbital configuration of the progenitors of DNSs. 
We find that the offset distributions of bound and unbound DNSs sensitively depend on the investigated physical ingredients. In particular, for tight NS+(pre-SN)He binaries, the mass ejection during the SN can produce large space velocities, even if the momentum kick imparted to the newborn NS is small. In addition to the gravity of the host galaxy, the star formation history can also strongly shape the expected offset distributions by altering the fractions of DNS of different merger times. For the GRB events occurring at high redshifts, the bound and unbound population are not distinguishable any more, as the merger time is limited by the local age of the Universe.

These combinations of physical ingredients introduce substantial degeneracies when interpreting observed offsets. Therefore, an observed offset alone can hardly constrain the properties of the DNS progenitor or the likelihood of a proposed host galaxy. A more robust host galaxy determination should combine the offset information with the chance-coincidence probability and independent constraints on the proposed host.

\begin{acknowledgments}
The authors thank the referee for the constructive
feedback on the manuscript.
X.-T.X. was supported by the Tsung-Dao Lee Postdoctoral fellowship at the Tsung-Dao Lee Institute and the National Natural Science Foundation of China (grant 12603055). B.-B.Z. acknowledges the support by the National Natural Science Foundation of China (grants 12573046 and 12121003), the Fundamental Research Funds for the Central Universities, and the Programme for Innovative Talents and Entrepreneurs in Jiangsu. 
\end{acknowledgments}

\bibliography{Xu_SMC}
\bibliographystyle{aasjournalv7}

\appendix

\section{Galactic potentials\label{app:isothermal}}

We describe here the four galaxy potential models implemented in
\texttt{GRB-offset}. All four models share a common parametrisation: the
host galaxy is characterised by its total virial mass, $M_{\rm vir}$, and the
virial radius,
\begin{equation}
    R_{\rm vir} = \left(\frac{3\,M_{\rm vir}}{4\pi\,\Delta_{\rm vir}
    \,\rho_{\rm crit}}\right)^{1/3},
    \label{eq:rvir}
\end{equation}
where $\Delta_{\rm vir}=200$ is the virial overdensity and
$\rho_{\rm crit}=3H_0^2/(8\pi G)$ is the critical density evaluated with
$H_0=70\,\kms\,{\rm Mpc}^{-1}$. For the reference model with
$M_{\rm vir}=10^{11}\,\mso$, Eq.\,\eqref{eq:rvir} gives
$R_{\rm vir}\approx95\,\kpc$.

\subsection{Isothermal\label{app:sub:iso}}

The truncated isothermal sphere is adopted as the reference potential
(Sect.\,\ref{sec:popsyn}). Its density profile follows $\rho\propto r^{-2}$,
which produces a mass enclosed within radius $r$ of
\begin{equation}
    M_{\rm iso}(r) = \begin{cases}
        M_{\rm vir}\,\dfrac{r}{R_{\rm vir}} & r \leq R_{\rm vir},\\[6pt]
        M_{\rm vir} & r > R_{\rm vir}.
    \end{cases}
    \label{eq:m_iso}
\end{equation}
The key consequence of this linear mass growth is a circular speed that is
constant everywhere inside the virial radius,
\begin{equation}
    v_c = \sqrt{\frac{G\,M_{\rm vir}}{R_{\rm vir}}},
    \label{eq:vc_iso}
\end{equation}
which gives $v_c\approx67\,\kms$ for the reference parameters.
The corresponding gravitational potential is
\begin{equation}
    \Phi_{\rm iso}(r) = \begin{cases}
        v_c^2\!\left[\ln\!\left(\dfrac{r}{R_{\rm vir}}\right)-1\right]
            & r \leq R_{\rm vir},\\[8pt]
        -\dfrac{G\,M_{\rm vir}}{r} & r > R_{\rm vir}.
    \end{cases}
    \label{eq:phi_iso}
\end{equation}
The flat rotation curve described by Eqs.\,\eqref{eq:m_iso}--\eqref{eq:phi_iso}
is a standard approximation for a dark-matter-dominated galaxy halo, and has
the computational advantage that $v_c$ and $R_{\rm vir}$ are the only free
parameters, both determined uniquely by $M_{\rm vir}$ through
Eqs.\,\eqref{eq:rvir} and \eqref{eq:vc_iso}.

\subsection{Pseudo-isothermal\label{app:sub:piso}}

The pseudo-isothermal sphere generalises the isothermal model by replacing
the central density cusp with a finite-density core of radius $r_c$:
\begin{equation}
    \rho_{\rm piso}(r) = \frac{\rho_0}{1+(r/r_c)^2}.
    \label{eq:rho_piso}
\end{equation}
At $r\ll r_c$ the density approaches the constant central value $\rho_0$,
while at $r\gg r_c$ the profile recovers the $r^{-2}$ behaviour of the
isothermal model. The enclosed mass is
\begin{equation}
    M_{\rm piso}(r) = 4\pi\,\rho_0\,r_c^2\!\left(r - r_c\arctan
    \frac{r}{r_c}\right), \quad r \leq R_{\rm vir},
    \label{eq:m_piso}
\end{equation}
with $\rho_0$ derived by imposing $M_{\rm piso}(R_{\rm vir})=M_{\rm vir}$. We
adopt a default core radius of $r_c=1\,\kpc$.

The gravitational potential admits an analytical closed form. Defining the
auxiliary function
\begin{equation}
    H(x) = \frac{\arctan x}{x} + \frac{1}{2}\ln(1+x^2),
    \label{eq:H_piso}
\end{equation}
with $H(0)=1$ (by L'H\^{o}pital's rule), the potential inside $R_{\rm vir}$
reads
\begin{equation}
    \Phi_{\rm piso}(r) = -\frac{G\,M_{\rm vir}}{R_{\rm vir}}
    - 4\pi G\,\rho_0\,r_c^2\left[H\!\left(\frac{R_{\rm vir}}{r_c}\right)
    -H\!\left(\frac{r}{r_c}\right)\right], \quad r \leq R_{\rm vir},
    \label{eq:phi_piso}
\end{equation}
and $\Phi_{\rm piso}(r)=-G\,M_{\rm vir}/r$ outside $R_{\rm vir}$.
The finite core softens the central potential relative to the isothermal
model, making the pseudo-isothermal sphere marginally less effective at
retaining newborn DNSs near $r_{\rm birth}$, which explains why the
Pseudo-isothermal model yields bound fractions and offset distributions very
similar to, but slightly lower than, those of the reference model
(Sect.\,\ref{res:param-study}).

\subsection{Milky Way-like\label{app:sub:mw}}

To represent a morphologically disk-like galaxy, we adopt an axisymmetric
three-component potential modelled after the structure of the Milky Way
\citep{Bovy2015ApJS..216...29B}. Using cylindrical coordinates $(R,\,z)$,
where $R$ is the galactocentric radius in the disk plane and $z$ the height
above it, the total potential is
\begin{equation}
    \Phi_{\rm MW}(R,z) = \Phi_{\rm bulge}(r) + \Phi_{\rm disk}(R,z)
    + \Phi_{\rm halo}(r),
    \label{eq:phi_mw}
\end{equation}
where $r=\sqrt{R^2+z^2}$ is the spherical galactocentric radius.

The stellar bulge is described by a Hernquist sphere
\citep{Hernquist1990ApJ...356..359H},
\begin{equation}
    \Phi_{\rm bulge}(r) = -\frac{G\,M_{\rm bulge}}{r+r_{\rm bulge}},
    \label{eq:phi_bulge}
\end{equation}
with a scale radius $r_{\rm bulge}$. The stellar disk follows the
Miyamoto-Nagai potential \citep{Miyamoto1975PASJ...27..533M},
\begin{equation}
    \Phi_{\rm disk}(R,z) = -\frac{G\,M_{\rm disk}}{\sqrt{R^2+
    \left(a_{\rm disk}+\sqrt{z^2+b_{\rm disk}^2}\right)^2}},
    \label{eq:phi_disk}
\end{equation}
where $a_{\rm disk}$ and $b_{\rm disk}$ are the disk scale length and scale
height. The dark matter halo is modelled by a Navarro-Frenk-White (NFW)
profile \citep{Navarro1996ApJ...462..563N},
\begin{equation}
    \Phi_{\rm halo}(r) = -\frac{G\,M_{\rm halo}}{f(c_{\rm halo})}
    \frac{\ln(1+r/r_s)}{r},
    \label{eq:phi_nfw}
\end{equation}
with scale radius $r_s=R_{\rm vir}/c_{\rm halo}$, concentration parameter
$c_{\rm halo}$, and the normalisation function $f(c)=\ln(1+c)-c/(1+c)$
that ensures $M_{\rm halo}(R_{\rm vir})=M_{\rm halo}$ exactly.

The component masses are set as fractions of $M_{\rm vir}$: $M_{\rm
bulge}=f_{\rm bulge}\,M_{\rm vir}$, $M_{\rm disk}=f_{\rm disk}\,M_{\rm vir}$,
and $M_{\rm halo}=f_{\rm halo}\,M_{\rm vir}$ with
$f_{\rm halo}=1-f_{\rm bulge}-f_{\rm disk}$. We adopt default values
$f_{\rm bulge}=0.05$, $r_{\rm bulge}=0.5\,\kpc$, $f_{\rm disk}=0.30$,
$a_{\rm disk}=3.0\,\kpc$, $b_{\rm disk}=0.28\,\kpc$, and $c_{\rm halo}=15$.
Compared with the isothermal and pseudo-isothermal models, this potential is
much more centrally concentrated: the combined bulge and disk account for 35\%
of the total mass within a few kpc, and the NFW halo profile rises steeply
towards the centre, dramatically increasing the local escape velocity at
$r_{\rm birth}$. As a result, very few newborn DNSs escape the host galaxy,
and the offset distribution of the bound population peaks near $r_{\rm birth}$
(Sect.\,\ref{res:param-study}). Because the disk component breaks spherical
symmetry, the offset for this model has to be computed by direct orbit integration
in \texttt{galpy}.

\subsection{Keplerian\label{app:sub:kep}}

The Keplerian potential treats the entire galaxy as a point mass at its
centre,
\begin{equation}
    \Phi_{\rm Kep}(r) = -\frac{G\,M_{\rm vir}}{r}.
    \label{eq:phi_kep}
\end{equation}
In this limit, the enclosed mass is $M(r)=M_{\rm vir}$ for all $r>0$, and
the circular speed decreases as
$\upsilon_{\rm circ}(r)=\sqrt{G\,M_{\rm vir}/r}$.
The Keplerian model is the most centrally concentrated of all four potentials.
Consequently, it
yields a unbound fraction of 12.7\% and the most compact offset distribution,
concentrated near $r_{\rm birth}$ (Sect.\,\ref{res:param-study}), making
it a useful upper-bound reference for the effect of central concentration.

\section{Further details of our Monte Carlo simulations\label{app:details}}

\subsection{Reference model}
\begin{figure}
    \centering
    \includegraphics[width=0.49\linewidth]{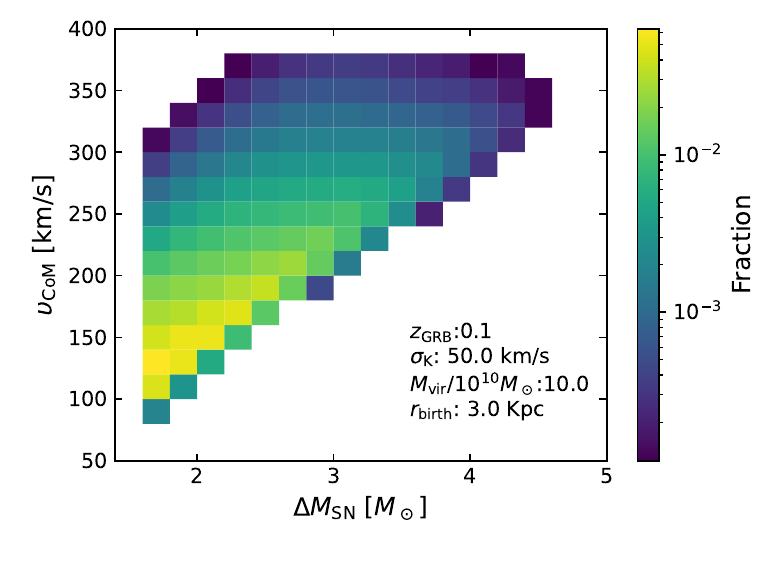}
    \includegraphics[width=0.49\linewidth]{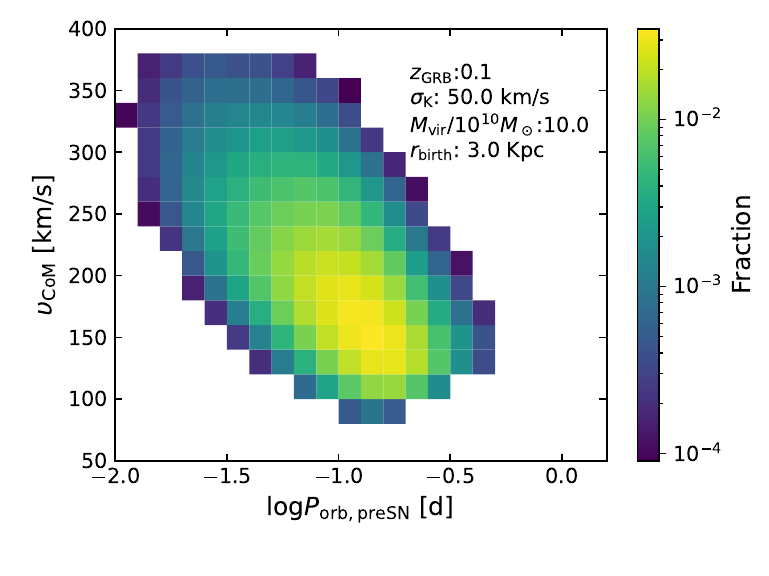}
    \vskip -0.7cm
    \caption{Correlation among the space velocity of the newborn DNS ($\upsilon_{\rm CoM}$), ejected mass during the second SN ($\Delta M_{\rm SN}=M_{\rm He}-M_{\rm NS}$), and the orbital periods of the NS+(pre-SN)He binaries ($\log P_{\rm orb,preSN}$). The calculated fraction in each pixel is colour-coded.}
    \label{fig:vcom-dm-pi}
\end{figure}

In Fig.\,\ref{fig:vcom-dm-pi}, we investigate the correlation among the space velocity of the newborn DNS ($\upsilon_{\rm CoM}$), ejected mass during the second SN ($\Delta M_{\rm SN}=M_{\rm He}-M_{\rm NS}$), and the orbital periods of the NS+(pre-SN)He binaries ($\log P_{\rm orb,preSN}$). As already discussed in the main text, the outcome $\upsilon_{\rm CoM}$ is dominated by the $\upsilon_{\rm rel}$ term in Eq.\,\eqref{eq:vcom}, which is
\begin{equation}
    \frac{\Delta M_{\rm SN}}{M_{\rm NS,2nd}+M_{\rm NS,1st}} \frac{M_{\rm NS,1st}}{M_{\rm He}+M_{\rm NS,1st}}\boldsymbol{\upsilon}_{\rm rel},
\end{equation}
where ``1st" and ``2nd" indicate the first- and second-born NSs. Here the relative velocity, $\upsilon_{\rm rel}$, is anti-correlated with $P_{\rm orb,preSN}$ (i.e., Eq.\,\ref{eq:vper0}). Therefore, $\upsilon_{\rm CoM}$ is expected to be correlated with $\Delta M_{\rm SN}$ (left panel in Fig.\,\ref{fig:vcom-dm-pi}) but anti-correlated with $\log P_{\rm orb,preSN}$ (right panel in Fig.\,\ref{fig:vcom-dm-pi}).

\subsection{Star formation history}

\begin{figure}
    \centering
    \includegraphics[width=\linewidth]{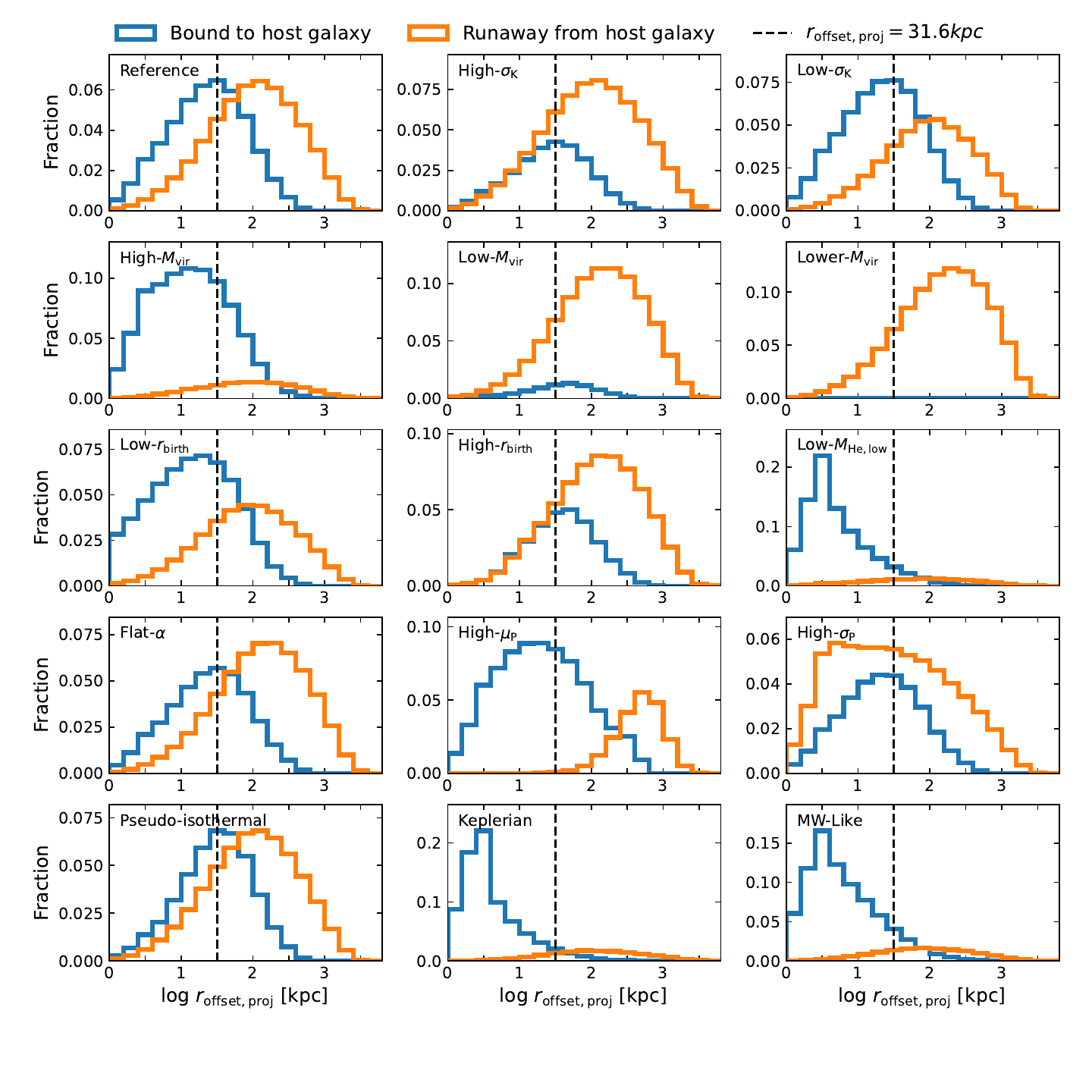}
        \vskip -1.5cm
    \caption{The same as Fig.\,\ref{fig:param-study-offset} but for a constant star formation rate.}
    \label{fig:param-study-constSFR}
\end{figure}

\begin{figure}
    \centering
    \includegraphics[width=\linewidth]{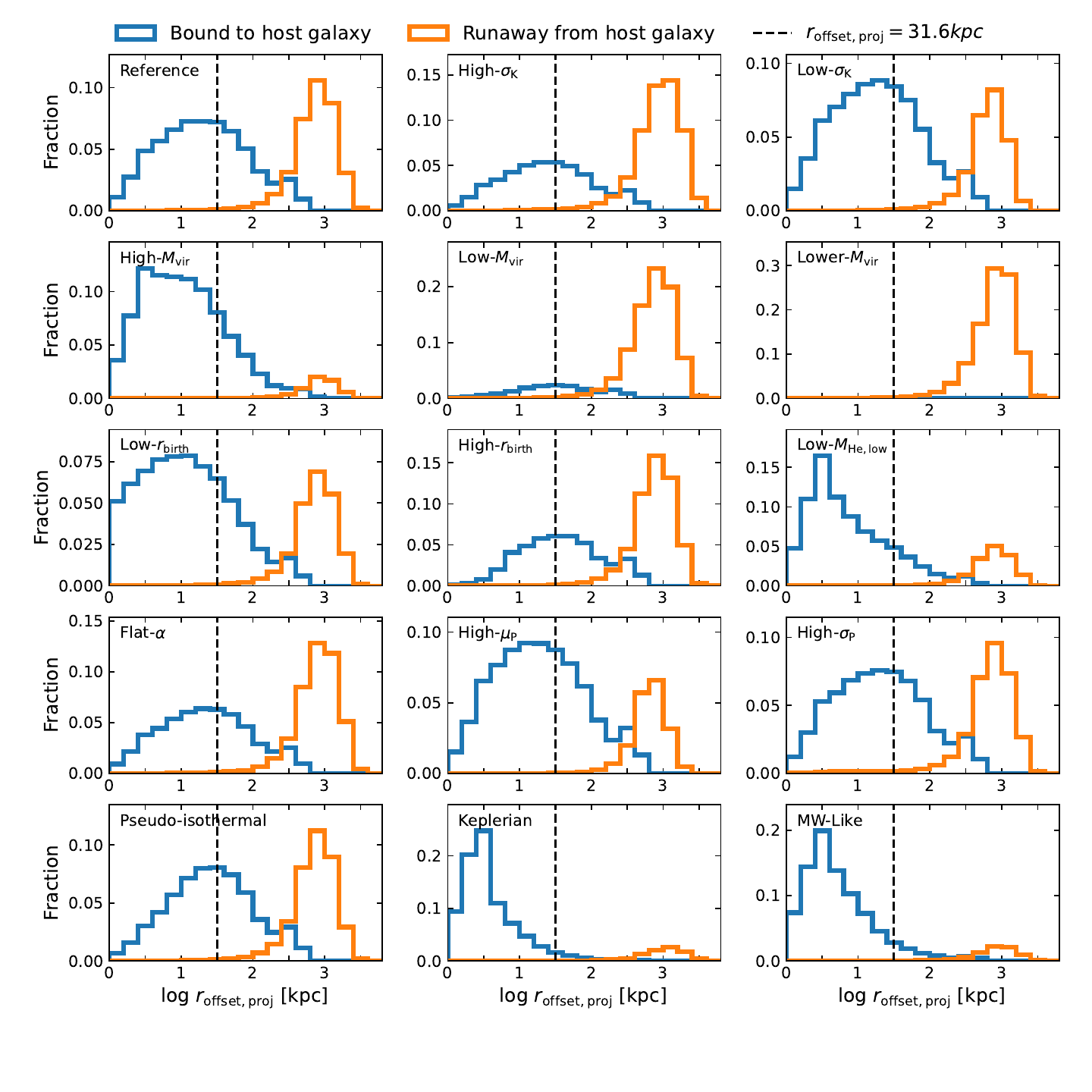}
        \vskip -1.5cm
    \caption{The same as Fig.\,\ref{fig:param-study-offset} but for the delayed-$\tau$ star formation history.}
    \label{fig:param-study-delayedtau}
\end{figure}

In this section, we repeat the parameter study in Sect.\,\ref{res:param-study} with a constant star formation rate (Fig.\,\ref{fig:param-study-constSFR}) and the delayed-$\tau$ star formation history (Fig.\,\ref{fig:param-study-delayedtau}). 

\end{CJK*}
\end{document}